\documentclass[aps, prd, article]{revtex4-2}

\usepackage{
	amsmath, 
	amssymb, 
	graphicx,
	hyperref, 
	float, 
	enumerate, 
	pgffor,
    longtable,
	multirow,
	mathrsfs,
	bm,
	slashed,
	xcolor
}

\def\bb#1{\left[#1\right]}
\def\cc#1{\left\{#1\right\}}

\def\CC#1{\Big\{#1\Big\}}

\def\pmat#1{\begin{pmatrix}#1\end{pmatrix}}

\def\d#1{\mathop{{\rm d}#1}}
\def\D#1#2{\mathop{{\rm d}^#1 #2}}

\DeclareMathOperator{\sgn}{sgn}

\def\lag{\mathscr{L}}

\def\del{\partial}

\def\vec#1{\bm{#1}}

\def\Ceu{C_{eu}}
\def\Ced{C_{ed}}
\def\Clqone{C_{\ell q}^{(1)}}
\def\Clqthree{C_{\ell q}^{(3)}}
\def\Clu{C_{\ell u}}
\def\Cld{C_{\ell d}}
\def\Cqe{C_{qe}}

\allowdisplaybreaks

\def\GeV{{\rm GeV}}
\def\TeV{{\rm TeV}}
\def\fb{{\rm fb}}

\DeclareMathAlphabet{\mymathbb}{U}{BOONDOX-ds}{m}{n}

\DeclareMathOperator{\Range}{\rm Range}

\begin{document}

\title{SMEFT probes in future precision DIS experiments}
\author{
    Chiara~Bissolotti${}^1$,
    Radja~Boughezal${}^1$, 
    Kaan~Simsek${}^{1,2}$
    \\ 
    \vspace{0.1cm}
    {\sl ${}^1$ Argonne National Laboratory, Lemont, IL, USA} \\
    {\sl ${}^2$ Northwestern University, Evanston, IL, USA} \\ 
}

\begin{abstract}
	We analyze the potential of future high-energy deep-inelastic scattering (DIS) experiments to probe new physics within the framework of the Standard Model Effective Field Theory (SMEFT). We perform a detailed study of SMEFT probes at a future Large Hadron-electron Collider (LHeC) and a Future Circular lepton-hadron Collider (FCC-eh) machine, and extend previous simulations of the potential of a Electron-Ion Collider (EIC) to include Z-boson vertex corrections. Precision Z-pole constraints on vertex corrections suffer from numerous degeneracies in the Wilson-coefficient parameter space. We find that both the LHeC and the FCC-eh can help remove these degeneracies present in the existing global fits of precision Z-pole observables and LHC data. The FCC-eh and LHeC will in many cases improve upon the existing precision electroweak bounds on the SMEFT parameter space. This highlights the important role of precision DIS measurements for new physics studies.
\end{abstract}

\maketitle
\tableofcontents

\section{Introduction\label{sec:intro}}
The accomplishments of the Standard Model (SM) are many. With the discovery of the Higgs boson in 2012, the predicted particle spectrum in the SM is complete.  However, the SM suffers from several shortcomings. The dark matter observed in the universe is not contained in the SM, nor are the mechanisms responsible for the baryon-antibaryon asymmetry and neutrino masses. Moreover, the SM contains numerous aesthetic issues, such as the electroweak hierarchy problem and the extreme hierarchy between fermion Yukawa couplings. A more complete and compelling theory is desirable. However, there has so far been neither conclusive evidence for new particles beyond the SM (BSM) nor any definitive deviation from SM predictions.

In an attempt to address these lingering issues in our understanding of Nature, many experiments have been launched or are under design. In this work, we consider the BSM potential of several proposed future electron-proton/deuteron deep-inelastic scattering (DIS) experiments: the Large Hadron-electron Collider (LHeC) \cite{LHeCStudyGroup:2012zhm}, the Future Circular lepton-hadron Collider (FCC-eh)~\cite{FCC:2018byv} and the Electron-Ion Collider (EIC) \cite{Accardi:2012qut}. The LHeC is a proposed upgrade of the Large Hadron Collider (LHC). It would operate alongside the LHC in order to utilize the LHC proton and ion beams. The earliest realistic operational period is estimated to be 2032, which coincides with the LHC Run 5 period. The integrated luminosity of the LHeC is projected to be of the order of $100~\fb^{-1}$. It will operate at center-of-mass (CM) energies reaching 1.5 TeV. It is designed to provide novel measurements in QCD, investigate DIS physics at low Bjorken-$x$ values, improve upon existing electroweak (EW) physics measurements, and probe BSM physics. The FCC-eh would occur at a new accelerator complex at CERN, and would feature center-of-mass energies approaching 3.5 TeV and integrated luminosities in the inverse attobarns~\cite{FCC:2018byv}. Like the LHeC it will feature a broad physics program spanning QCD and electroweak measurements to new physics searches. The EIC is a United States Department of Energy project that will be constructed at Brookhaven National Laboratory (BNL). The EIC will be the first high-energy DIS machine that collides polarized electrons with polarized protons. It is anticipated to commence operating within a decade. The EIC is designed to collide a polarized electron beam of energy 5 to 18 GeV with polarized proton beams of energies 41 to 275 GeV, with polarized light ions of energies up to 166 GeV, and with unpolarized heavy ions of energies up to 110 GeV. It will run at CM energies between fixed-target-scattering and high-energy colliders, namely 70 to 140 GeV. It will improve the extraction of parity-violating (PV) DIS asymmetries in EW neutral-current (NC) cross section with reduced uncertainties from luminosity and detector acceptance/efficiency.

Our goal in this work is to study the BSM potential of the LHeC, FCC-eh and the EIC  with a detailed accounting of anticipated uncertainties. We consider the neutral-current (NC) DIS cross section as our observable at the LHeC, following previous studies of electroweak physics at the LHeC~\cite{Britzger:2020kgg} and FCC-eh~\cite{Britzger:2022abi}. At the EIC we focus on PV asymmetries, following earlier studies of BSM physics at the EIC~\cite{Boughezal:2022pmb}. Since there has been no conclusive sign of new particles beyond the SM yet, we perform our analysis within the framework of the Standard Model Effective Field Theory (SMEFT) (see \cite{Brivio:2017vri} for a review of the SMEFT). In the SMEFT, one builds higher-dimensional operators using the existing SM particle spectrum. All new physics is assumed to be heavier than the SM states, as well as the accessible collider energies. The leading order basis of the SMEFT for on-shell fields has been completely classified up to dimension-12 \cite{Buchmuller:1985jz, Arzt:1994gp, Grzadkowski:2010es, Murphy:2020rsh, Li:2020gnx, Harlander:2023psl}. In this work, we restrict ourselves to dimension-6 (there is a lepton-number violating operator at dimension-5, which is irrelevant to our study). Previous work has shown that DIS measurements at the EIC and in low-energy fixed target experiments can resolve blind spots in the semi-leptonic four fermion Wilson coefficient space that remain after Drell-Yan measurements at the LHC~\cite{Boughezal:2020uwq, Boughezal:2021kla}, and that EIC measurements of single-spin asymmetries can competitively probe Wilson coefficients of dipole operators~\cite{Boughezal:2023ooo}. We consider here the full spectrum of Wilson coefficients that can alter the DIS process at leading-order in the SMEFT loop expansion. These include both semi-leptonic four fermion Wilson coefficients and $Z$-boson vertex correction factors, for a total of 17 Wilson coefficients at leading order in the SMEFT loop expansion. It is traditionally assumed that the vertex corrections are best measured with $Z$-pole precision EW observables at LEP and SLC. However, due to the limited kinematic information available from $Z$-pole data there are numerous degeneracies between the Wilson coefficients. This was illustrated for example in~\cite{Ellis:2020unq}, where in addition contributions from existing LHC data were also considered. The obtained constraints on the Wilson coefficients can become an order of magnitude weaker when the full spectrum of Wilson coefficients is activated, as compared to turning on only a single coefficient. We show here that the future DIS measurements can help resolve these degeneracies. This and other key aspects of our study are summarized below. 
\begin{itemize}

\item We find that the LHeC and FCC-eh can significantly extend the search reach for semi-leptonic four fermion operators. While the EIC can probe the SMEFT operators of interest to a few TeV, the LHeC and FCC-eh can exceed 10 TeV. We thoroughly study different beam energy, polarization options, luminosity assumptions and lepton species choices at all three colliders. We find that no single choice probes the entire SMEFT parameter space, and that a full spectrum of  run scenarios is needed to fully explore the physics possibilities beyond the SM.

\item We find that the option of a positron beam in future DIS experiments can significantly extend there reaches in certain sectors of the Wilson coefficient parameter space, due to the structure of the underlying matrix elements.

\item It is often assumed that the most stringent constraints on universal shifts of the $Z$-boson vertex couplings to fermions are obtained from fits to the precision $Z$-pole observables. While this is true when only a single Wilson coefficient is turned on, when several are activated simultaneously numerous degeneracies arise, as demonstrated in~\cite{Ellis:2020unq}. We show that the LHeC and FCC-eh can improve upon the existing bounds on the $Z$-boson 
couplings by resolving these degeneracies.

\end{itemize}

This manuscript is organized as follows. In Section \ref{sec:review}, we briefly review the relevant formalism of the SMEFT and DIS and also define our observables of interest. In Section \ref{sec:analysis},  we describe our LHeC, FCC-eh and EIC pseudodata sets, as well as anticipated uncertainties. We also detail our numerical procedure for fitting of the SMEFT parameters. In Section \ref{sec:results}, we present the results of the fits. We conclude in Section \ref{sec:conclusion}. Details regarding the construction of the error matrix and the generation of pseudodata are given in the Appendix.

\section{Review of the formalism\label{sec:review}}
\subsection{Review of the SMEFT formalism}

The SMEFT is a model-independent extension of the SM Lagrangian in which one builds operators of dimension higher than four, $O_k^{(n)}$, using the existing spectrum of the SM and assuming the SM gauge symmetries. The Wilson coefficients associated with these operators are denoted as $C_k^{(n)}$.  These effective couplings are defined at a UV cut-off scale, $\Lambda$, which is assumed to be heavier than all the SM states and all accessible collider energies. The Lagrangian takes the form
\begin{align}
    \lag_{\rm SMEFT} = 
        \lag_{\rm SM}
        + \sum_{n > 4} {1 \over \Lambda^{n-4}} \sum_k C_k^{(n)} O_k^{(n)}
\end{align}
In this work, we restrict ourselves to operators of dimension-6. There is only a single lepton-number violating operator at dimension 5, which is irrelevant to our analysis of lepton-number conserving observables. We note that we linearize our observables in the Wilson coefficients. We will see later that the results obtained justify this assumption. There are 17 operators that affect NC DIS matrix elements at leading order in coupling constants \cite{Grzadkowski:2010es}, which are summarized in Table \ref{tab:ops}.
\begin{table}
    [H]
    \centering
    \caption{Dimension-6 operators in the Warsaw basis \cite{Grzadkowski:2010es} affecting NC DIS matrix elements at leading order in the coupling constants. Operators in the left column shift the $ffV$ vertices, while those on the right induce semi-leptonic four-fermion contact interactions. Both the operators and their associated Wilson coefficients are shown.}
    \label{tab:ops}
    \begin{tabular}{|c|c|c|c|}
        \hline 
        \multicolumn{2}{|c|}{$ffV$} & \multicolumn{2}{|c|}{semi-leptonic four-fermion} \\
        \hline 
        $C_{\varphi WB}$ & $O_{\varphi WB} = (\varphi^\dagger \tau^I \varphi) W_{\mu\nu}^I B^{\mu\nu}$ & $\Clqone$ & $O_{\ell q}^{(1)} = (\bar \ell \gamma_\mu \ell) (\bar q \gamma^\mu q)$ \\ 
        \hline 
        $C_{\varphi D}$ & $O_{\varphi D} = (\varphi^\dagger D_\mu \varphi)^* (\varphi^\dagger D^\mu \varphi)$ & $\Clqthree$ & $(\bar \ell \gamma_\mu \tau^I \ell) (\bar q \gamma^\mu \tau^I q)$ \\
        \hline 
        $C_{\varphi \ell}^{(1)}$ & $O_{\varphi \ell}^{(1)} = (\varphi^\dagger i \stackrel{\leftrightarrow}{D}_\mu \varphi) (\bar \ell \gamma^\mu \ell)$ & $\Ceu$ & $O_{eu} = (\bar e \gamma_\mu e) (\bar u \gamma^\mu u)$ \\
        \hline 
        $C_{\varphi \ell}^{(3)}$ & $O_{\varphi \ell}^{(3)} = (\varphi ^\dagger i \stackrel{\leftrightarrow}{D}_\mu \tau^I \varphi) (\bar \ell \gamma^\mu \tau^I \ell)$ & $\Ced$ & $O_{ed} = (\bar e \gamma_\mu e) (\bar d \gamma^\mu d)$ \\ 
        \hline 
        $C_{\varphi e}$ & $O_{\varphi e} = (\varphi^\dagger i \stackrel{\leftrightarrow}{D}_\mu \varphi) (\bar e \gamma^\mu e)$ & $\Clu$ & $O_{\ell u} = (\bar \ell \gamma_\mu \ell) (\bar u \gamma^\mu u)$ \\
        \hline 
        $C_{\varphi q}^{(1)}$ & $O_{\varphi q}^{(1)} = (\varphi^\dagger i \stackrel{\leftrightarrow}{D}_\mu \varphi) (\bar q \gamma^\mu q)$ & $\Cld$ & $O_{\ell d} = (\bar \ell \gamma_\mu \ell) (\bar d \gamma^\mu d)$\\ 
        \hline 
        $C_{\varphi q}^{(3)}$ & $O_{\varphi q}^{(3)} = (\varphi ^\dagger i \stackrel{\leftrightarrow}{D}_\mu \tau^I \varphi) (\bar q \gamma^\mu \tau^I q)$ & $\Cqe$ & $O_{qe} = (\bar q \gamma_\mu q) (\bar e \gamma^\mu e)$ \\ 
        \hline 
        $C_{\varphi u}$ & $O_{\varphi u} = (\varphi^\dagger i \stackrel{\leftrightarrow}{D}_\mu \varphi) (\bar u \gamma^\mu u)$ \\ 
        \cline{1-2}
        $C_{\varphi d}$ & $O_{\varphi d} = (\varphi^\dagger i \stackrel{\leftrightarrow}{D}_\mu \varphi) (\bar d \gamma^\mu d)$  \\
        \cline{1-2}
        $C_{\ell \ell}$ & $O_{\ell \ell} = (\bar \ell \gamma_\mu \ell) (\bar \ell \gamma^\mu \ell)$  \\
        \cline{1-2} 
        \cline{1-2}
    \end{tabular} 
\end{table}
Here, $\varphi$ is the SU(2) Higgs doublet, $\ell$ and $q$ are the left-handed lepton and quark doublets, and $e$, $u$, and $d$ are the right-handed electron and up- and down-quark singlets, respectively. The $\tau^I$ are the Pauli matrices and the double-arrow covariant derivative is defined such that
\begin{gather}
    \varphi^\dagger i \stackrel{\leftrightarrow}{D}_\mu \varphi = \varphi^\dagger i D_\mu \varphi + {\rm h.c.} \\ 
    \varphi^\dagger i \stackrel{\leftrightarrow}{D}_\mu \tau^I \varphi = \varphi^\dagger i D_\mu \tau^I \varphi + {\rm h.c.}
\end{gather}
We suppress flavor indices and assume flavor universality in our analysis for simplicity. We remark that operators containing scalar and dipole fermionic bilinears are discarded in our analysis. Such vertex factors produce cross section contributions proportional to fermion masses, which are small and are  neglected here. We note that SMEFT loop corrections are expected to be sub-dominant to the next-to-leading order (NLO) QCD corrections. Since we include the NLO QCD corrections in our study and find that they do not have a large effect on the obtained results, we assume that the higher-order terms in the SMEFT loop expansion can be safely neglected.

\subsection{Review of the DIS formalism}

\begin{figure}
    [H]
    \centering
    \includegraphics[height=3cm]{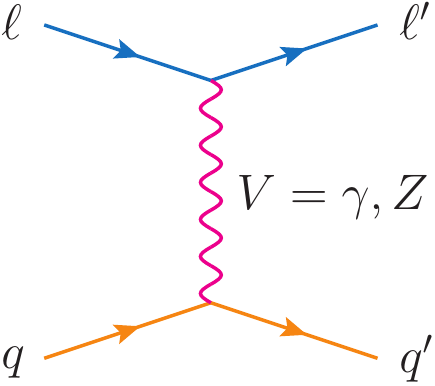}
    \hspace{3cm}
    \includegraphics[height=3cm]{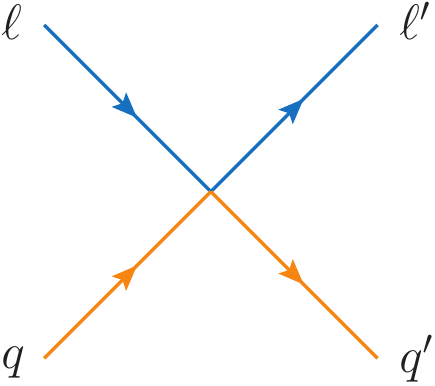}
    \caption{LO Feynman diagrams for the partonic process mediating $\ell + H \to \ell' + X$.}
    \label{fig:lo-feynman}
\end{figure}

\begin{figure}
    [htbp]
    \centering
    \includegraphics[height=3cm]{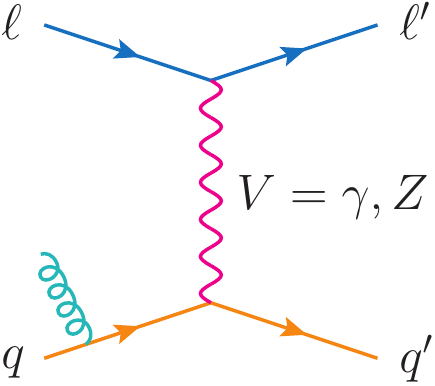}
    \hfill
    \includegraphics[height=3cm]{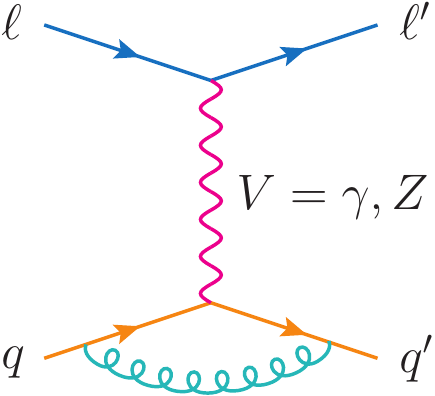}
    \hfill
    \includegraphics[height=3cm]{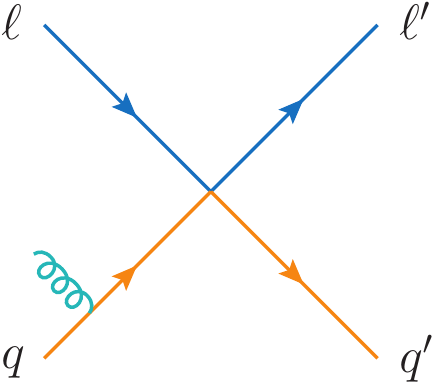}
    \hfill
    \includegraphics[height=3cm]{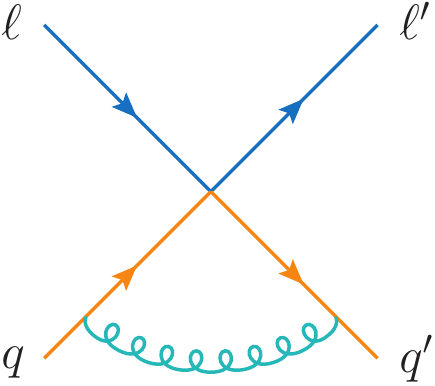}
    \caption{Representative Feynman diagrams describing the NLO QCD corrections to $\ell + H \to \ell' + X$.}
    \label{fig:nlo-feynman}
\end{figure}

In our analysis we study NC DIS in the process $\ell + H \to \ell' + X$, where $\ell$ is an electron or a positron, $H$ can be a proton or a deuteron, and $\ell'$ and $X$ are the final-state lepton and hadronic systems, respectively. Charged current DIS involves missing energy and consequently the reconstruction of hadronic final states in order to determine the kinematic variables. It therefore receives different and typically larger systematic uncertainties. For this reason we do not consider this mode in our study. We include next-to-leading order QCD corrections to both the SM and the SMEFT corrections. At leading order in the perturbative QCD expansion, this process can be mediated by single photon or $Z$-boson exchange or by the SMEFT contact interaction of leptons with quarks. The LO Feynman diagrams are presented in Figure \ref{fig:lo-feynman}.
The NLO QCD corrections to the SM process are well known \cite{deFlorian:2012wk, Altarelli:1979kv, Vogelsang:1990ug, Altarelli:1979ub, deFlorian:1994wp}. These corrections modify only the quark lines, as illustrated in Fig.~\ref{fig:nlo-feynman}. Therefore the corrections are identical for both SM and SMEFT cross sections. It is convenient to express the DIS cross sections in terms of structure functions. 

The NC DIS cross-section expressions for collisions of a lepton $\ell$ with an unpolarized or polarized hadron are given in terms of the NC structure functions $F_{1,3,L}^{\rm NC}$ and $g_{1,5,L}^{\rm NC}$ by
\begin{align}
    {\D2{\sigma^\ell_{\rm NC}} \over \d x \d {Q^2}} = 
        {2 \pi \alpha^2 \over x Q^4} \CC{
            [1+(1-y)^2] 2x F_1^{\rm NC}
            + \sgn(\ell) [1-(1-y)^2]x F_3^{\rm NC}
            + (1-y) 2 F_L^{\rm NC}
        }
\end{align}
and
\begin{align}
    {\D2{\Delta\sigma^\ell_{\rm NC}} \over \d x \d {Q^2}} = 
        {8 \pi \alpha^2 \over x Q^4} \CC{
            [1+(1-y)^2] xg_5^{\rm NC}
            - \sgn(\ell) [1-(1-y)^2] xg_1^{\rm NC}
            + (1-y) g_L^{\rm NC}
        }.
\end{align}
where $\sgn$ is the \textit{particle signum} function that returns $+1$ for particles and $-1$ for antiparticles. $Q$ is the usual DIS momentum transfer, $x$ is the Bjorken variable, and $y$ is the inelasticity parameter. These are defined as usual for the DIS process. 
We define the reduced cross sections as 
\begin{align}
    {\D2{\sigma_{r, {\rm NC}}^\ell} \over \d x \d {Q^2}} &= 
        \cc{
            {2 \pi \alpha^2 \over x Q^4} [1 + (1-y)^2]
        }^{-1} {\D2{\sigma^\ell_{\rm NC}} \over \d x \d {Q^2}}, \\ 
    {\D2{\Delta \sigma_{r, {\rm NC}}^\ell} \over \d x \d {Q^2}} &= 
        \cc{
            {4 \pi \alpha^2 \over x Q^4} [1 + (1-y)^2]
        }^{-1} {\D2{\Delta\sigma^\ell_{\rm NC}} \over \d x \d {Q^2}}.
\end{align}
From this point onward, when we mention cross sections, we mean the reduced ones and denote them simply by $(\Delta)\sigma_{\rm NC}$.

\begin{figure}
    [H]
    \centering
    \includegraphics[width=.45\textwidth]{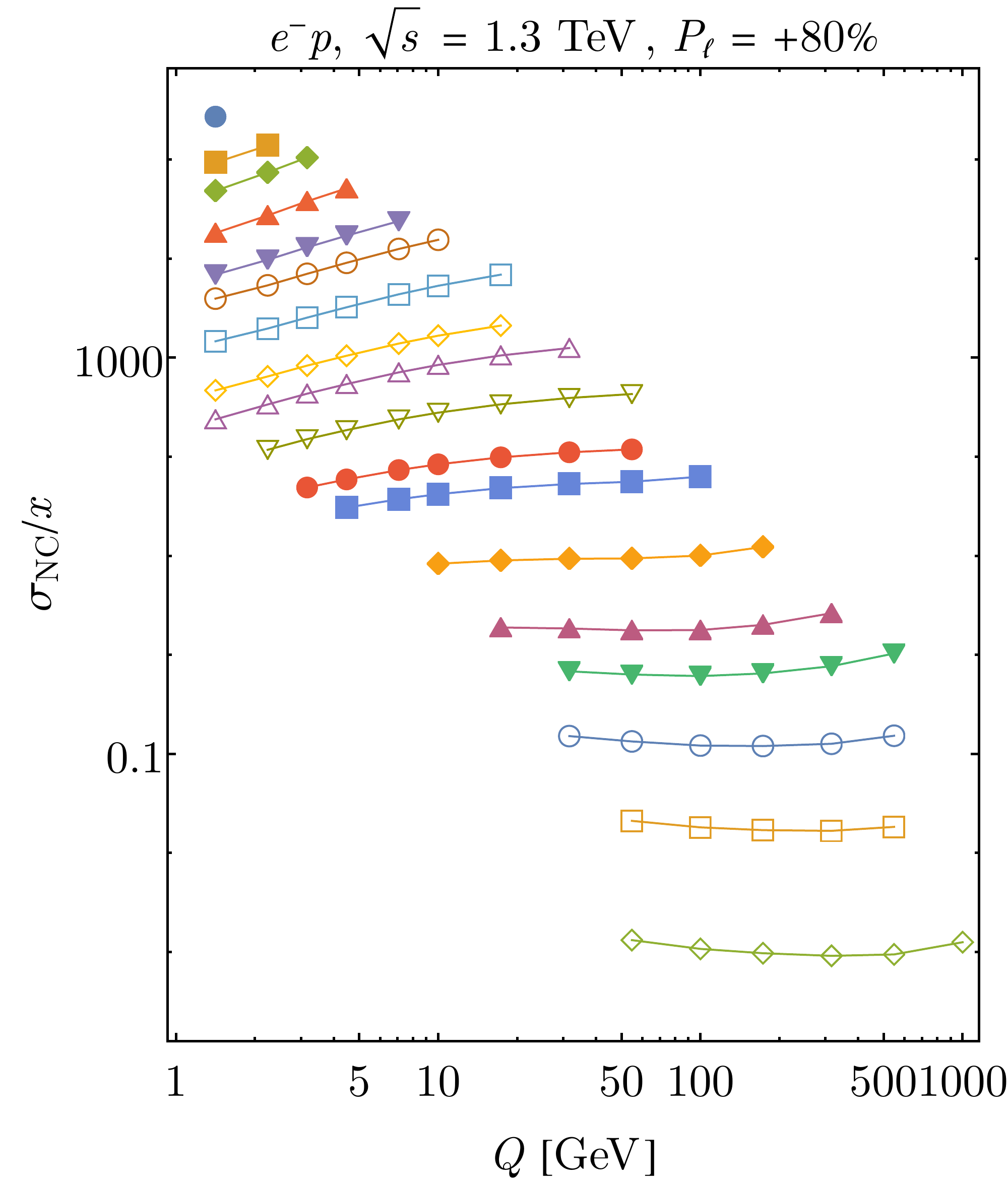}
    \includegraphics[width=.45\textwidth]{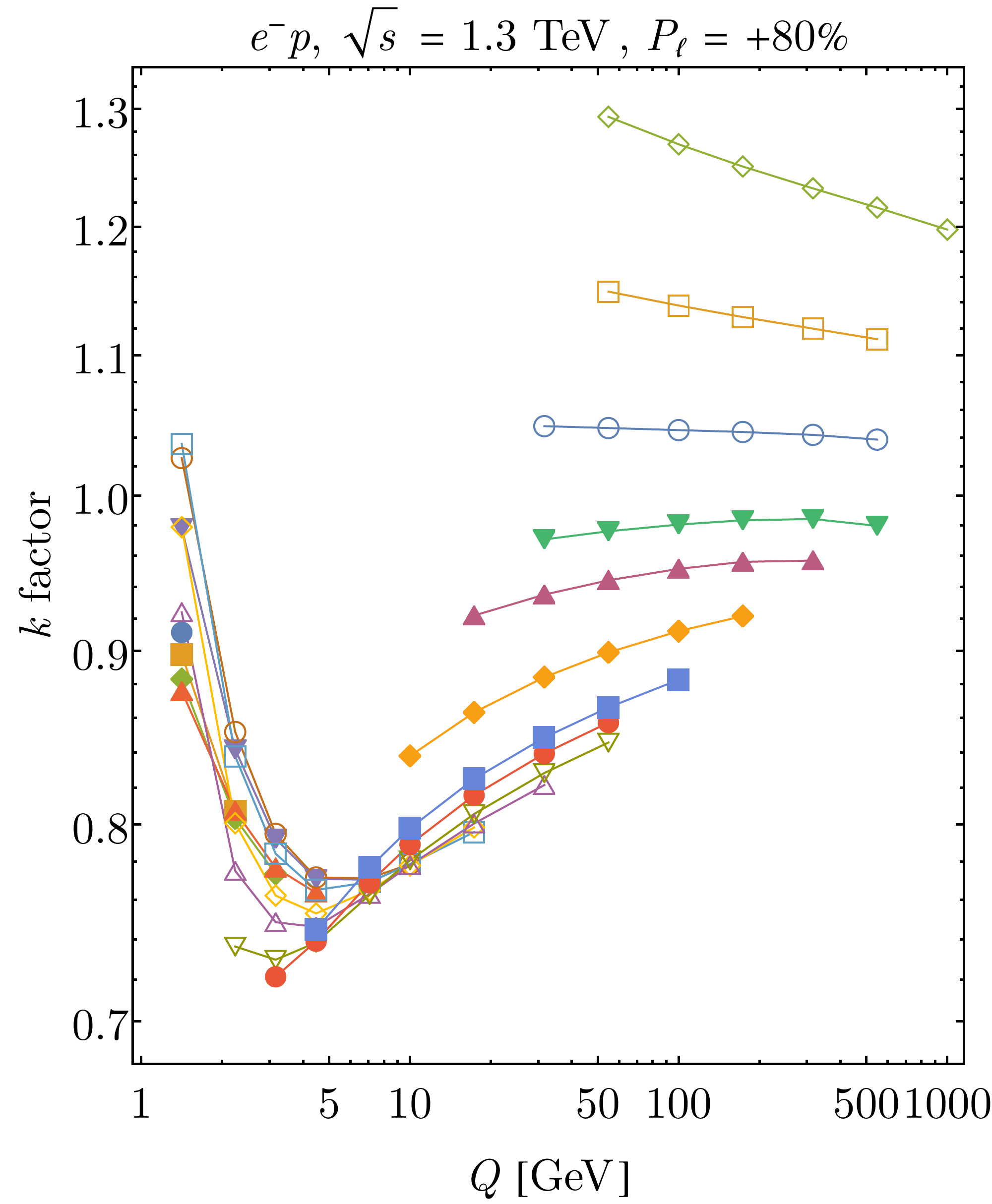}\\
    \vspace{.5cm}
    \includegraphics[width=.7\textwidth]{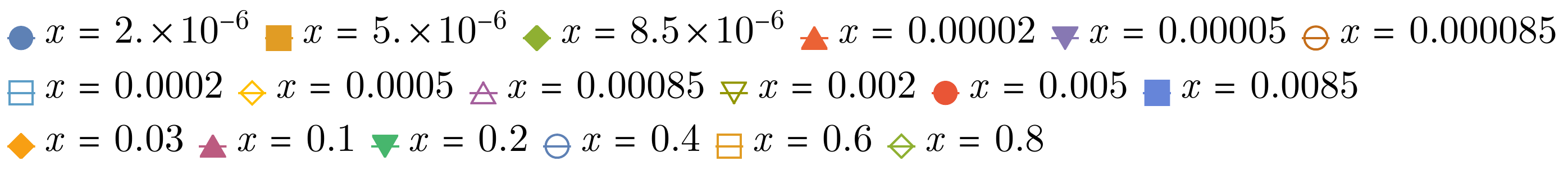}
    \caption{NC DIS cross section with NLO QCD corrections for $e^-p$ collisions at $\sqrt s = 1.3~\TeV$ with $P_\ell = +80\%$.}
    \label{fig:nlo-qcd-xsection-kfactor}
\end{figure}

In Fig.~\ref{fig:nlo-qcd-xsection-kfactor}, we show the NC DIS cross section with NLO QCD corrections for $e^-p$ collisions at $\sqrt s = 1.3~\TeV$ with right-handed (RH) electrons of polarization $P_\ell = +80\%$ and the corresponding $k$ factors as a function of $Q$ for various $x$ values. We observe that the NLO QCD corrections to the NC DIS cross section are 30\% at most. They exhibit high sensitivity to $Q$ and low sensitivity to $x$ for $Q \lesssim 30~\GeV$, and low sensitivity to $Q$ and high sensitivity to $x$ for higher values of $Q$. We have compared the Wilson coefficient constraints obtained using NLO QCD structure functions with those obtained using LO QCD, and have found very similar results. We therefore believe that the neglect of QCD corrections at the NNLO level and beyond are justified in our analysis.

\subsection{Observables of interest}

The observable of interest at the LHeC and FCC-eh is the NC DIS cross section, $\sigma_{\rm NC}$, of unpolarized protons/deuterons with electrons or positrons of various polarizations. We choose this observable in order to compare our simulated pseudodata with previous studies in the literature~\cite{Britzger:2020kgg,Britzger:2022abi}. For the EIC we consider PV asymmetries in cross sections of polarized electrons with either polarized or unpolarized protons/deuterons. Previous studies have shown that this asymmetry at the EIC generically provides somewhat more sensitivity to BSM effects than asymmetries with polarized protons or lepton-charge asymmetries~\cite{Boughezal:2022pmb}, so we focus on this case here. We define the unpolarized PV asymmetry by
\begin{align}
    A_{\rm PV} =
        {
        \sigma_{\rm NC}^+ - \sigma_{\rm NC}^- \over 
        \sigma_{\rm NC}^+ + \sigma_{\rm NC}^- 
        }.
\end{align}
and the polarized one by
\begin{align}
    \Delta A_{\rm PV} =
    {  
       \Delta \sigma_{\rm NC}^0 \over \sigma_{\rm NC}^0 .
    }
\end{align}
Here, $\sigma_{\rm NC}^\pm$ is the unpolarized NC DIS $e^-H$ ($H=p,D$) cross section evaluated with $\lambda_\ell = \pm P_\ell$, $\sigma_{\rm NC}^0$ is the same as $\sigma_{\rm NC}^\pm$ but with $\lambda_\ell = 0$, and $\Delta \sigma_{\rm NC}^0$ is the same as $\sigma_{\rm NC}^0$ but with a polarized hadron. $P_\ell$ is the assumed value for the lepton beam polarization at the EIC. We linearize the SMEFT expressions in this study. Thus, the SMEFT observables have the generic form
\begin{align}
    \mathcal O = \mathcal O^{\rm SM} + \sum_k C_k \ \delta \mathcal O_k + \mathcal O (C_k^2)
\end{align}
where $k$ runs over the active Wilson coefficients, $\mathcal O = \sigma_{\rm NC}$ or $A_{\rm PV} $ is the observable, and $\delta \mathcal O_k$ is the SMEFT correction to the observable proportional to the Wilson coefficient $C_k$. 

\section{Description of the analysis\label{sec:analysis}}
\subsection{Description of the pseudodata}
\label{sec:pseudodata}

\begin{figure}
    [htbp]
    \centering
    \includegraphics[height=.35\textheight]{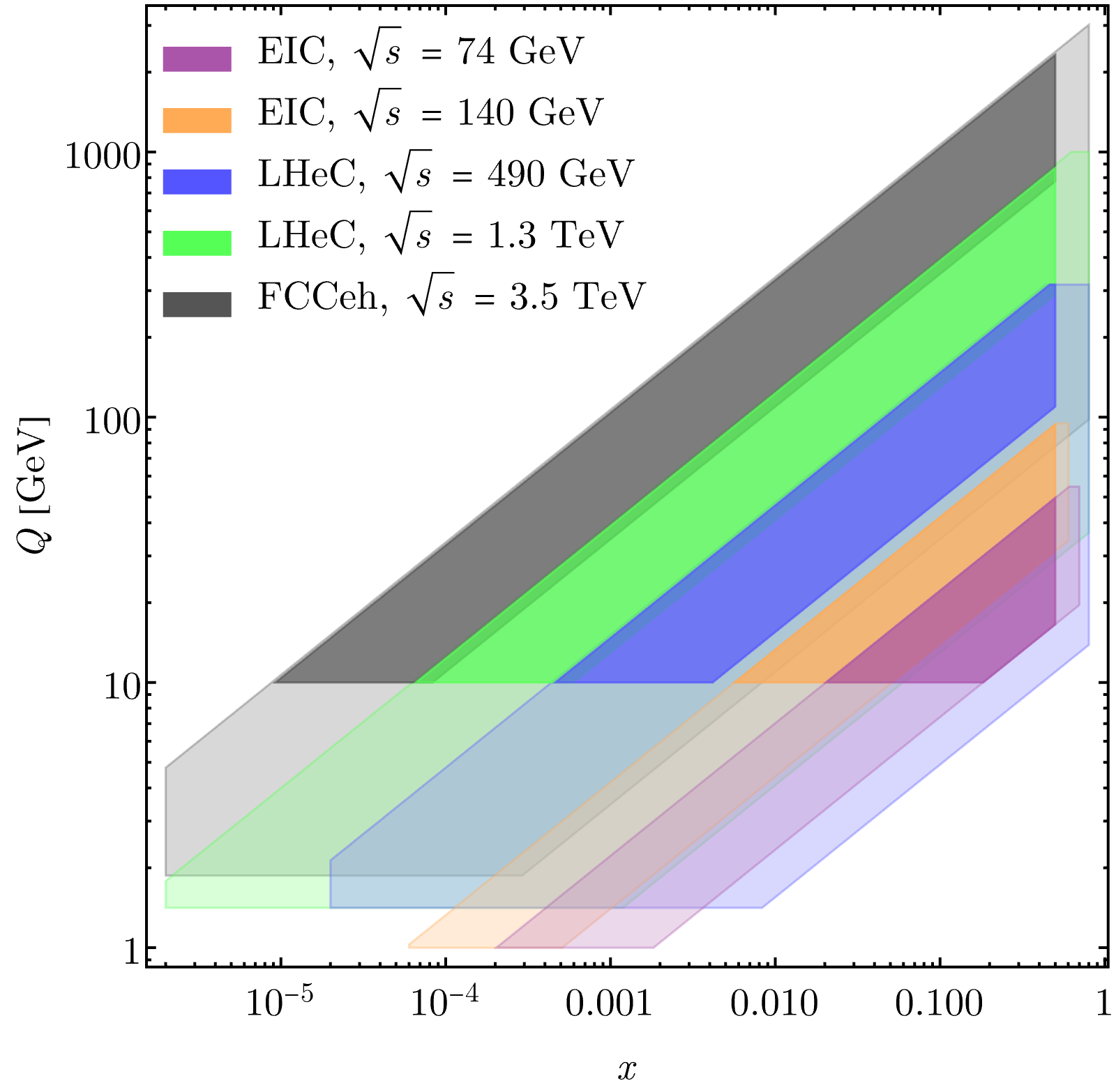}
    \caption{Kinematic coverage of LHeC, FCC-eh and EIC data sets for several choices of center-of-mass energy.}
    \label{fig:kinematic-coverage}
\end{figure}

For our analysis we use the most recent publicly available LHeC pseudodata sets \cite{r:KleinData, r:KleinPaper}, as well as the EIC data set found to be most sensitive to SMEFT Wilson coefficients in \cite{Boughezal:2022pmb}. For the FCC-eh we generate pseudodata sets following the procedure established in~\cite{Boughezal:2022pmb}, with the FCC-eh run parameters found in~\cite{Britzger:2022abi}. We refer to the pseudodata sets as \textit{data sets} from this point onward. In Table \ref{tab:data-sets}, we summarize the configurations of these data sets in terms of beam energies, lepton beam polarizations, and total integrated luminosities, together with our labeling scheme and also the observable of interest. Note that in this work, we do not consider a possible 10-fold-high-luminosity scenario of the EIC. We also consider joint LHeC and FCC-eh fits that combine all run scenarios for each experiment listed below.
\begin{table}
    [htbp]
    \centering
    \caption{Configuration of the LHeC, FCC-eh and EIC data sets used in our analysis in terms of beam energies, lepton and hadron beam polarizations, total integrated luminosities, our labeling scheme, and the observable of interest.}
    \label{tab:data-sets}
    \begin{tabular}{|c|c|l|c|}
        \hline 
        Experiment & Data set label & Data set configuration & Observable \\
        \hline 
        \hline 
        \multirow{7}{*}{LHeC} & LHeC1 & $60~\GeV \times 1000~\GeV\ e^-p,\ P_\ell = 0,\ \mathcal L = 100~\fb^{-1}$ & \multirow{7}{*}{$\sigma_{\rm NC}$} \\ 
        \cline{2-3}
        & LHeC2 & $60~\GeV \times 7000~\GeV\ e^-p,\ P_\ell = -80\%,\ \mathcal L = 100~\fb^{-1}$ & \\
        \cline{2-3}
        & LHeC3 & $60~\GeV \times 7000~\GeV\ e^-p,\ P_\ell = +80\%,\ \mathcal L = 30~\fb^{-1}$ & \\
        \cline{2-3}
        & LHeC4 & $60~\GeV \times 7000~\GeV\ e^+p,\ P_\ell = +80\%,\ \mathcal L = 10~\fb^{-1}$ & \\
        \cline{2-3}
        & LHeC5 & $60~\GeV \times 7000~\GeV\ e^-p,\ P_\ell = -80\%,\ \mathcal L = 1000~\fb^{-1}$ & \\
        \cline{2-3}
        & LHeC6 & $60~\GeV \times 7000~\GeV\ e^-p,\ P_\ell = +80\%,\ \mathcal L = 300~\fb^{-1}$ & \\
        \cline{2-3}
        & LHeC7 & $60~\GeV \times 7000~\GeV\ e^+p,\ P_\ell = 0\%,\ \mathcal L = 100~\fb^{-1}$ & \\
        \hline 
        \hline        
        \multirow{3}{*}{FCC-eh} & FCCeh1 & $60~{\rm GeV} \times 50000~{\rm GeV}~e^-p,~P_\ell = -80\%,~\mathcal L = 2~{\rm ab}^{-1}$ & \multirow{3}{*}{$\sigma_{\rm NC}$} \\ 
\cline{2-3}
& FCCeh2 & $60~{\rm GeV} \times 50000~{\rm GeV}~e^-p,~P_\ell = +80\%,~\mathcal L = 0.5~{\rm ab}^{-1}$ & \\ 
\cline{2-3}
& FCCeh3 & $60~{\rm GeV} \times 50000~{\rm GeV}~e^+p,~P_\ell = 0,~\mathcal L = 0.2~{\rm ab}^{-1}$ & \\ 
\hline
\hline
        \multirow{8}{*}{EIC} & D4 & $10~\GeV \times 137~\GeV \ e^-D, \ P_\ell = 80\%, \ \mathcal L = 100~\fb^{-1}$ & \multirow{4}{*}{$A_{\rm PV}$}  \\
        \cline{2-3}
        & D5 & $18~\GeV \times 137~\GeV \ e^-D, \ P_\ell = 80\%, \ \mathcal L = 15.4~\fb^{-1}$ & \\
        \cline{2-3}
        & P4 & $10~\GeV \times 275~\GeV \ e^-p, \ P_\ell = 80\%, \ \mathcal L = 100~\fb^{-1}$ & \\
        \cline{2-3}
        & P5 & $18~\GeV \times 275~\GeV \ e^-p, \ P_\ell = 80\%, \ \mathcal L = 15.4~\fb^{-1}$ & \\
        \cline{2-4}
        & $\Delta$D4 & The same as D4 but with $P_\ell = 0$ and $P_H = 70\%$ & \multirow{4}{*}{$\Delta A_{\rm PV}$} \\
        \cline{2-3}
        & $\Delta$D5 & The same as D5 but with $P_\ell = 0$ and $P_H = 70\%$ & \\
        \cline{2-3}
        & $\Delta$P4 & The same as P4 but with $P_\ell = 0$ and $P_H = 70\%$ & \\
        \cline{2-3}
        & $\Delta$P5 & The same as P5 but with $P_\ell = 0$ and $P_H = 70\%$ & \\
        \cline{2-4}

        \hline 
    \end{tabular}  
\end{table}
We restrict ourselves to bins that satisfy $x\leq0.5$, $Q\geq10~\GeV$, and $0.1\leq y\leq 0.9$. We introduce these additional cuts to avoid large uncertainties from nonperturbative QCD and nuclear dynamics that occur at low $Q$ and high $x$, where we expect SMEFT effects to be reduced. We call the bins that pass our cuts the \textit{good} bins. The kinematic coverage of the LHeC, FCC-eh and EIC data sets is displayed in Fig.~\ref{fig:kinematic-coverage}. The darker regions in these plots indicate the good regions. 

We next discuss the anticipated error budgets for these data sets. For the LHeC and FCC-eh, we use the error estimates provided in previous analyses~\cite{Britzger:2020kgg,Britzger:2022abi}. The uncertainty components consist of uncorrelated statistical uncertainties ($\delta \sigma_{\rm stat}$), uncorrelated efficiency errors ($\delta\sigma_{\rm ueff}$), and correlated systematic errors ($\delta\sigma_{\rm sys}$). The correlated systematic uncertainties include contributions from lepton energy scale and polar angle measurements ($\delta \sigma_{\rm len}$ and $\delta\sigma_{\rm lpol}$), the hadronic energy scale ($\delta\sigma_{\rm hen}$), radiative corrections ($\delta\sigma_{\rm rad}$), photoproduction backgrounds ($\delta\sigma_{\rm gam}$), a global efficiency factor ($\delta \sigma_{\rm geff}$), and luminosity ($\delta\sigma_{\rm lum}$). We assume the luminosity error to be 1\% relative to the cross section. We introduce the systematics in a fully correlated manner. As for the EIC asymmetries, we have statistical uncertainties given by event counts, corrected for beam polarization and lepton beam luminosities:
$\delta A_{\rm PV, stat} = {1 \over P_\ell \sqrt N}$, $\delta\Delta A_{\rm PV, stat} = {P_\ell \over P_H} \ \delta A_{\rm PV, stat}$, where $P_\ell$ is the assumed lepton beam polarization at the EIC and $P_H$ is the assumed proton/deuteron polarization. The assumed systematic errors $\delta A_{\rm PV, sys}$ are mainly due to particle background and other imperfections in measurements. They are assumed to be uncorrelated and 1\% relative to the asymmetry. We assume uncertainties in lepton (hadron) beam polarization, $\delta(\Delta)A_{\rm pol}$, to be fully correlated and 1\% (2\%) relative in asymmetry. More discussion on the anticipated experimental uncertainties at the EIC is given in~\cite{Boughezal:2022pmb}. Additionally, for all data sets we take into account PDF errors fully correlated between bins, $\delta\sigma_{\rm pdf}$ and $\delta A_{\rm PV, pdf}$, respectively. We summarize the expected uncertainties for both the EIC, FCC-eh and LHeC in Table~\ref{tab:uncertainties} below. In Appendix~\ref{app:errmat} we discuss how these systematic uncertainties are incorporated into the error matrix for our analysis. We also give details of our pseudodata generation and describe our statistical procedure for deriving Wilson coefficients bounds in Appendix~\ref{app:errmat}.

 \begin{table}
    [htbp]
    \centering
    \caption{Anticipated values or ranges of uncertainties at the LHeC, FCC-eh and the EIC for the \textit{good} bins used in our analysis. All uncertainties are relative with respect to the observable of interest, unless stated otherwise.}
    \label{tab:uncertainties}
    \begin{tabular}{|c|l|c|c|}
        \hline 
        Experiment & Source of uncertainty & Value or range of uncertainty [\%] & Observable \\
        \hline 
        \hline 
        \multirow{10}{*}{LHeC} & Statistical & 0.10--6.83 &\multirow{10}{*}{$\sigma_{\rm NC}$} \\
        \cline{2-3}
        & Uncorrelated efficiency & 0.50 &\\
        \cline{2-3}
        & Lepton energy & 0.11--0.49 &\\
        \cline{2-3}
        & Lepton polar angle & 0.00--0.13 &\\
        \cline{2-3}
        & Hadron energy & 0.00--1.81 &\\
        \cline{2-3}
        & Radiative corrections & 0.30 & \\
        \cline{2-3}
        & Photoproduction background & 0.00--1.00 & \\
        \cline{2-3}
        & Global efficiency & 0.50 & \\
        \cline{2-3}
        & Calorimetry noise & 0.00 & \\
        \cline{2-3}
        & Luminosity & 1.00 & \\
        \hline 
        \hline 
         \multirow{8}{*}{FCC-eh} & Statistical & 0.10--5.49 & \multirow{8}{*}{$\sigma_{\rm NC}$} \\ 
        \cline{2-3}
        & Lepton energy & 0.90 & \\
        \cline{2-3}
        & Lepton polar angle & 0.40 & \\ 
        \cline{2-3}
        & Hadron energy & 2.00 & \\ 
        \cline{2-3}
        & Radiative corrections & 0.30 & \\ 
        \cline{2-3}
        & Photoproduction background & 0.00--1.00 & \\ 
        \cline{2-3}
        & Global efficiency & 0.50 & \\ 
        \cline{2-3}
        & Luminosity & 1.00 & \\ 
        \hline
        \hline
        \multirow{10}{*}{EIC} & Statistical & 1.53--65.87 & \multirow{3}{*}{$A_{\rm PV}$}\\
        \cline{2-3}
        & Systematical & 1.00 & \\
        \cline{2-3}
        & Lepton beam polarization & 1.00 & \\
        \cline{2-4}
        & Statistical & 1.74--75.28 & \multirow{3}{*}{$\Delta A_{\rm PV}$} \\
        \cline{2-3}
        & Systematical & 1.00 & \\
        \cline{2-3}
        & Hadron beam polarization & 2.00 & \\
        \cline{2-4}
        & Statistical & 3.86--166.64 & \multirow{4}{*}{$A_{\rm LC}$} \\
        \cline{2-3}
        & Systematical & 1.00 & \\
        \cline{2-3}
        & Luminosity & 2.00 [absolute] & \\
        \cline{2-3}
        & NLO QED & 0.00--0.51 [absolute] & \\
       	        \hline 
    \end{tabular}  
 \end{table}

\begin{figure}
    [htbp]
    \centering
    \includegraphics[height=.27\textheight]{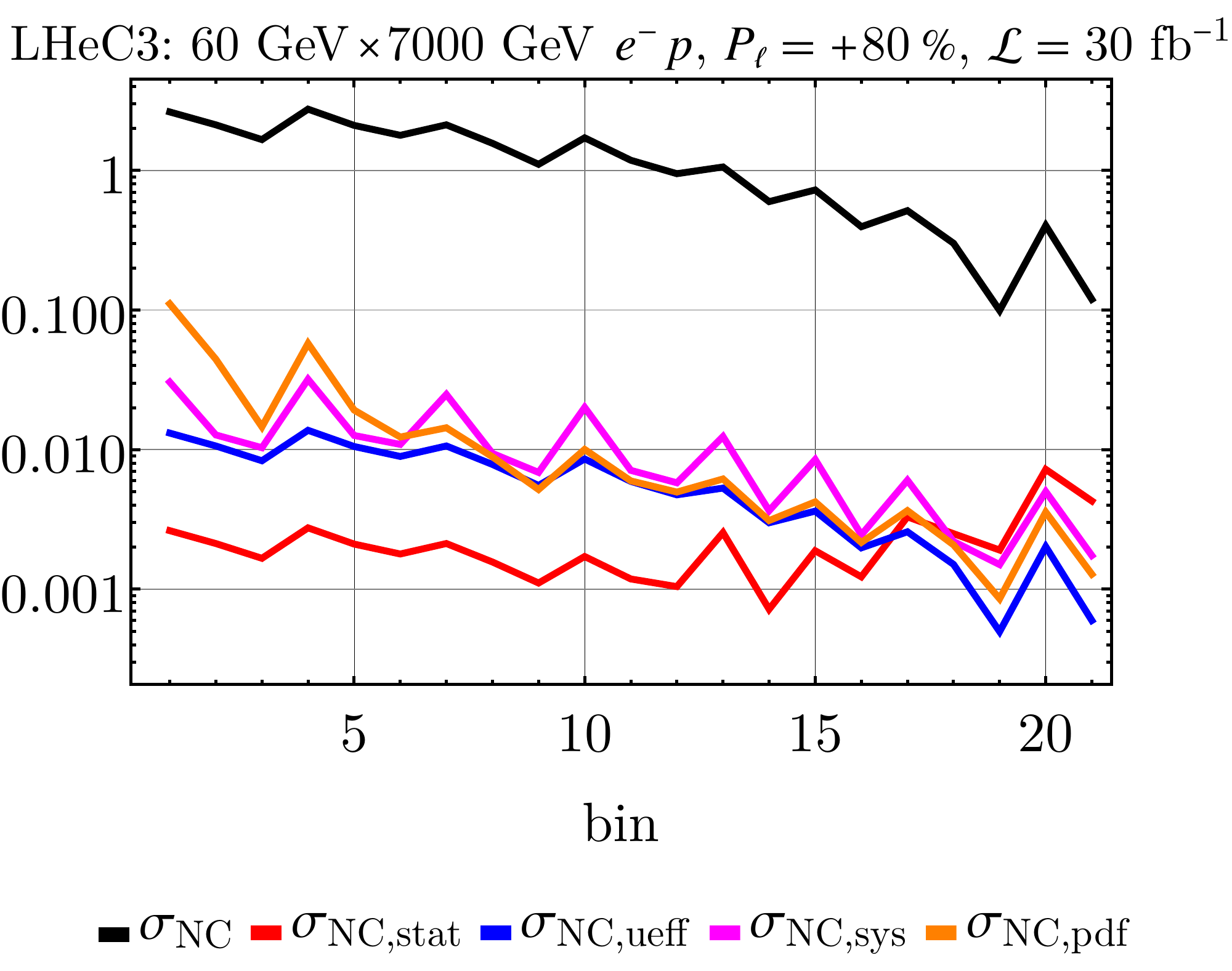}
    \includegraphics[height=.28\textheight]{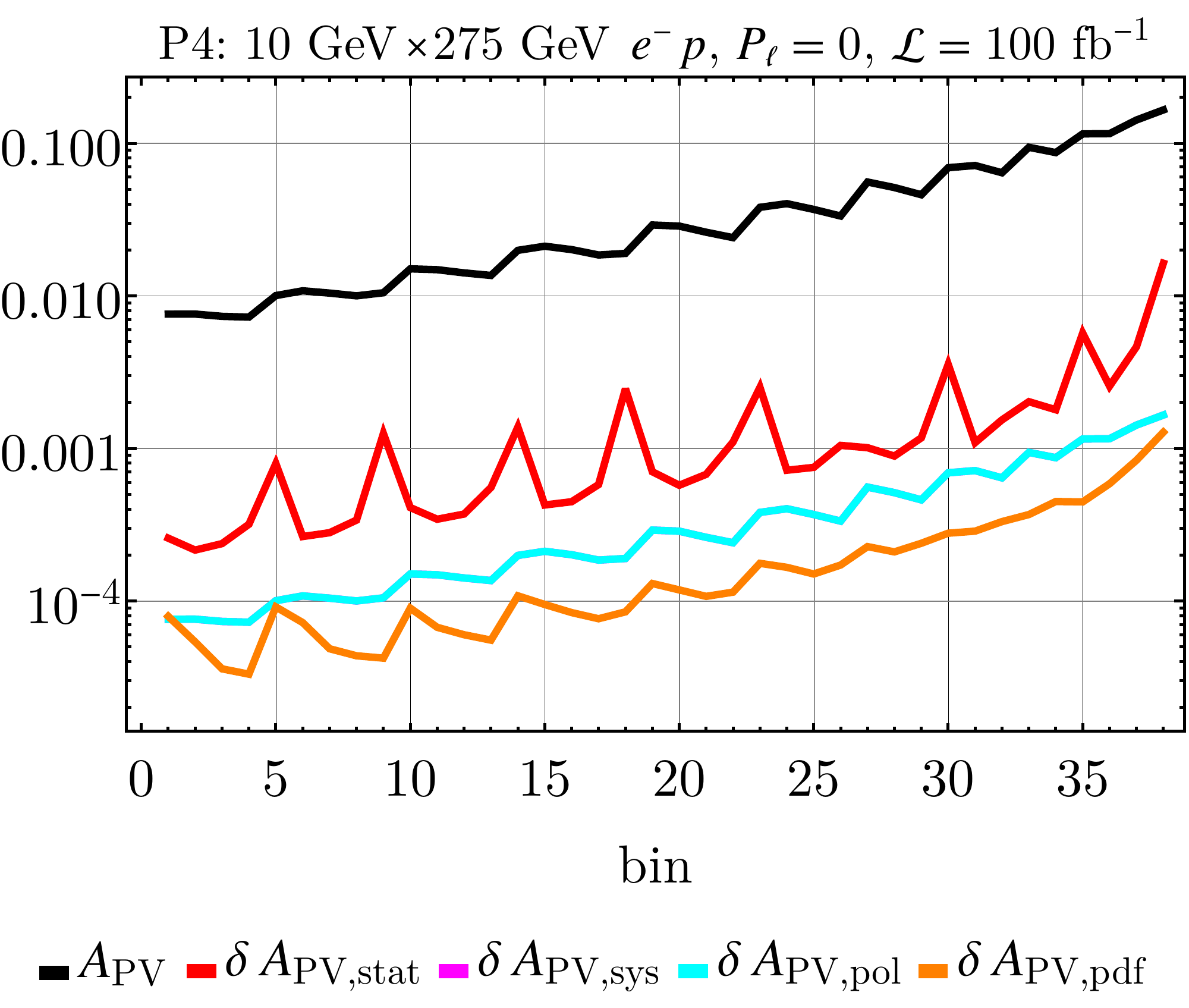}
    \caption{The various uncertainty components that enter the diagonal entries of the error matrix for data sets LHeC3 (left) and P4 (right). The red line denotes the statistical uncertainty, the blue line denotes the uncorrelated global efficiency uncertainty, the magenta line indicates the systematic uncertainty, and the orange line is the PDF uncertainty. For P4, the cyan line denotes the beam polarization uncertainty.}
    \label{fig:lhec-p4-uncertainties}
\end{figure}

\begin{figure}
    [htbp]
    \centering
    \includegraphics[height=.28\textheight]{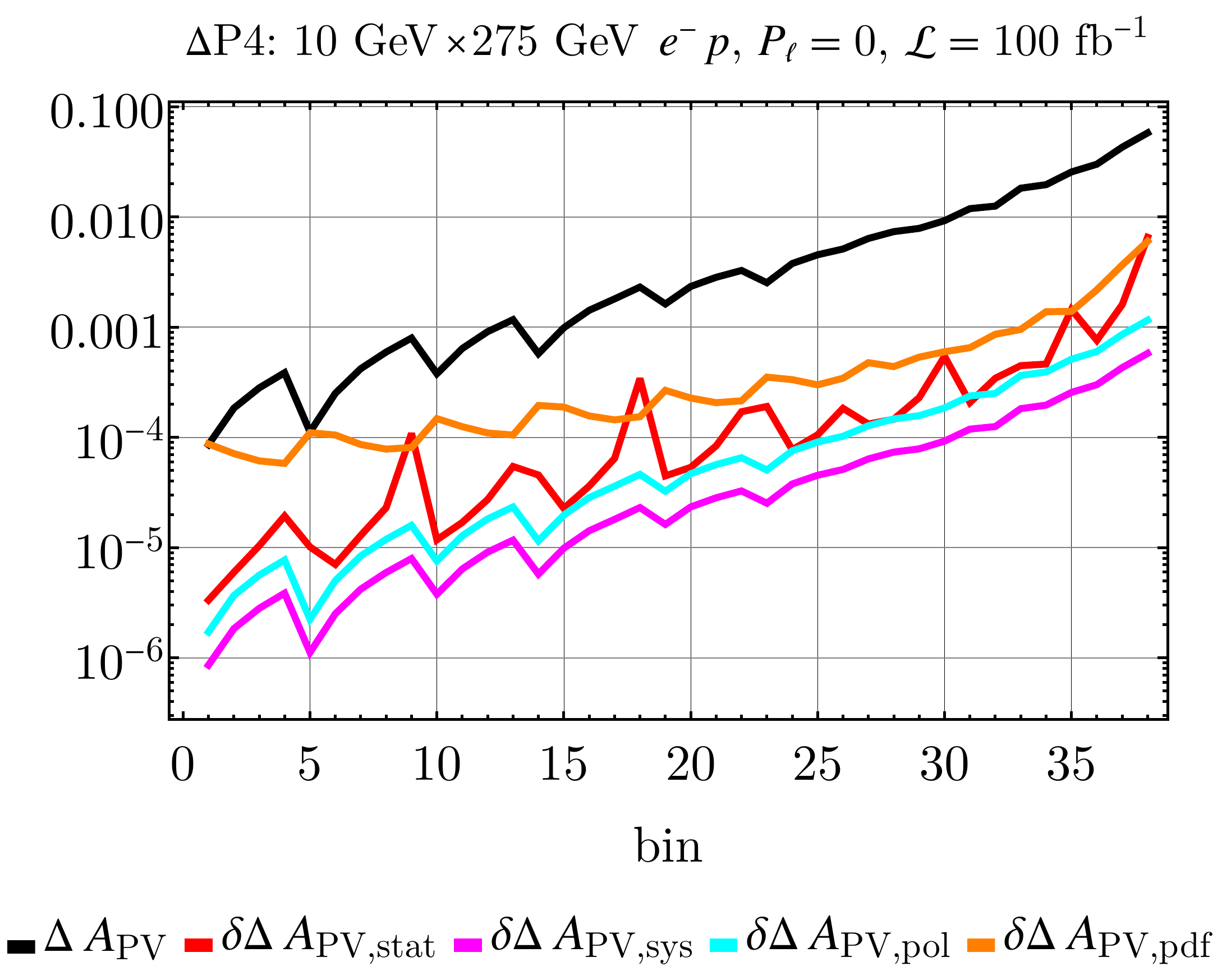}
     \includegraphics[height=.28\textheight]{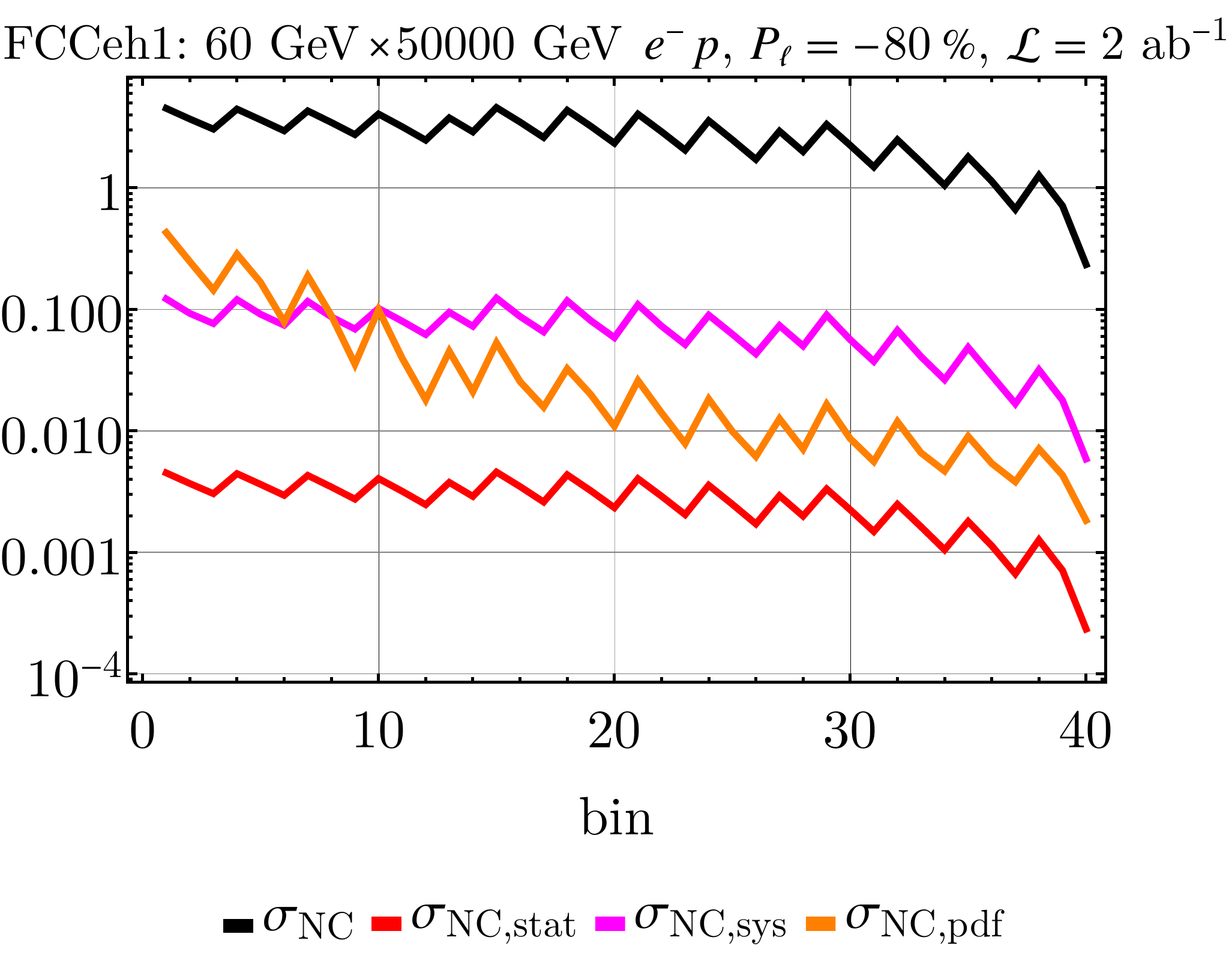}
    \caption{The same as in Figure \ref{fig:lhec-p4-uncertainties} but for $\Delta$P4 (left panel) and for the FCC-eh (right panel).}
    \label{fig:p4-deltap4-uncertainties}
\end{figure}

In Figs.~\ref{fig:lhec-p4-uncertainties} and~\ref{fig:p4-deltap4-uncertainties} we present the aforementioned uncertainty components at the LHeC, FCC-eh and EIC for representative data sets. On the horizontal axis, we order the bins of the indicated data sets. On the vertical axes are the central values of the observables and the uncertainty components that go into the diagonal entries of the error matrix. The central values of cross sections and asymmetries are denoted by the black lines, statistical uncertainties by red, systematics by magenta, and PDF errors by orange. The blue line for the LHeC indicates uncorrelated efficiency errors, whereas the cyan lines for P4 and $\Delta$ P4 denote beam polarization errors. The bins are sorted first by low to high $Q$ and then by low to high $x$. This explains the observed sawtooth behavior. 

From Figs.~\ref{fig:lhec-p4-uncertainties} and~\ref{fig:p4-deltap4-uncertainties} we can make the following points.
\begin{itemize}

    \item Statistical uncertainties are smaller for the LHeC and FCC-eh runs for the majority of the bins compared to the EIC. They constitute only a negligible part of the total uncertainties for these machines. Measurements at the LHeC and FCC-eh will be systematics limited. We note that the PDF error constitutes a non-negligible fraction of the error budget, indicating the need to eventually consider a joint fit of PDFs and Wilson coefficients~\cite{Carrazza:2019sec,Greljo:2021kvv,Gao:2022srd}. 
    
    \item  The statistical uncertainties are the leading error for the unpolarized PV asymmetries at the EIC, and are almost an order of magnitude larger than other sources of errors. For the polarized PV asymmetries, statistical uncertainties are smaller than the PDF errors. 
    
    \item Other uncorrelated uncertainties originating from efficiency errors compete with the systematics and PDF errors at the LHeC. Systematic uncertainties dominate for most of the FCC-eh bins.
        
    \item Correlated uncertainties at the LHeC account for the largest source of errors for the majority of the bins used in our analysis. At the EIC, for the PV asymmetries, the only correlated uncertainty comes from beam polarization, and it is a small part of the total uncertainty.
        
\end{itemize}

\section{SMEFT fit results\label{sec:results}}
We discuss here our numerical results. For our input parameters we use an electroweak scheme with $G_F$, $\alpha$, and $M_Z$ as our inputs. The numerical values for the parameters used in our analysis are as follows:
\begin{align}
    G_F &= 1.1663787 \times 10^{-5}~\GeV^{-2} \\
    \alpha^{-1} &= 137.036 \\ 
    M_Z &= 91.1876~\GeV .
\end{align}
We assume a lepton beam polarization at the EIC of $P_\ell = 80\%$, and a hadron beam polarization of $P_H=70\%$. The assumed polarizations for the various LHeC and FCC-eh runs are given in Table~\ref{tab:data-sets}.  For our UV scale we take $\Lambda = 1~\TeV$. We use {\tt NNPDF3.1} NLO PDFs \cite{NNPDF:2017mvq} for the unpolarized cross sections and {\tt NNPDF 1.1}  NLO polarized PDFs~\cite{Nocera:2014gqa}. The 2-loop running strong coupling constant is numerically evaluated according to the renormalization group equation
\begin{align}
    \mu_R^2 {\d {\alpha_s} \over \d{\mu_R^2}} = \beta(\alpha_s) = -(b_0 \alpha_s^2 + b_1 \alpha_s^3)
\end{align}
where $b_0 = {33-2N_f \over 12\pi}$ and $b_1 = {153 - 19N_f \over 24\pi^2}$ with the initial condition $\alpha_s(M_Z^2) = 0.1185$. We set $\mu_R^2 = Q^2$ and take $N_f = 5$ since we impose the cut $Q>10$ GeV on our data. 

\subsection{Bounds on semi-leptonic four-fermion operators}

We begin by activating only the seven semi-leptonic four-fermion operators. Previous studies have shown that the Drell-Yan process at the LHC, the natural channel to probe these operators due to its energy reach and excellent measurement precision, has difficulty probing certain linear combinations of Wilson coefficient in this subspace~\cite{Alte:2018xgc,Boughezal:2020uwq}. Future DIS experiments can help resolve these degeneracies~\cite{Boughezal:2020uwq,Boughezal:2022pmb}. Restricting ourselves to this subspace of Wilson coefficients allows us to compare the potential of DIS measurements at the EIC, FCC-eh and LHeC to improve upon Drell-Yan measurements at the LHC. The marginalized 95\% confidence level (CL) bounds on the semi-leptonic four-fermion Wilson coefficients projected from the full seven-parameter (7d) fit and the corresponding effective UV scales are presented in Table~\ref{tab:1d-bounds-uv-table}. We also present the bounds obtained by activating only single operators for comparison. We consider several different fit scenarios in this table: fits of the separate EIC data sets P4, $\Delta$P4, the combined EIC fit of D4, $\Delta$D4, P4, and $\Delta$P4, the individual LHeC runs, a joint LHeC fit, the individual FCC-eh fits, and a joint FCC-eh fit. We can make the following points from this table.
\begin{itemize}

\item There are significant differences between the marginalized and non-marginalized bounds in fits to individual data sets. When we activate the entire sector of four-fermion operators we observe strong correlations among Wilson coefficients, leading to degeneracies. However, when we combine the different run scenarios at a given machine the flat directions in the respective individual fits are removed. The effective scales probed in the fully marginalized joint fits range from 500 GeV to 1 TeV at the EIC, from 2.5 to 14 TeV at the LHeC, and from 2.0 to 18 TeV at the FCC-eh, depending on the Wilson coefficient being considered.

\item The polarized PV asymmetries at the EIC play an important role in the fully marginalized joint fit, even though they typically lead to weaker constraints on individual coefficients. 

 \item No single LHeC or FCC-eh data set can provide strong probes of all the four-fermion semi-leptonic Wilson coefficients. This is not surprising since the different runs each utilize distinct lepton helicities and species, and since these seven Wilson coefficients characterize the strength of lepton-quark contact interactions for different helicity states. The full spectra of proposed run scenarios at both the LHeC and FCC-eh are needed to fully explore the allowed parameter space. These possibilities are represented in the figure by the joint LHeC and FCC-eh bounds.
   
    \item The joint LHeC data set imposes significantly stronger bounds on semi-leptonic four-fermion Wilson coefficients than the EIC. This is also not surprising, given its higher momentum transfers where SMEFT-induced deviations are expected to be larger. For the majority of operators the joint FCC-eh fit imposes stronger constraints than the joint LHeC fit.
        
    \item $e^-p$ collisions with RH electrons (LHeC3, LHeC6 and FCCeh2) provide the optimal configurations to constrain $\Ceu$ and $\Ced$.
    
    \item The highest-luminosity $e^-p$ collisions with LH electrons (LHeC2, LHeC5 and FCCeh1) are the optimal configuration to constrain $\Clqone$ and $\Clqthree$.
    
    \item Polarized $e^+p$ collisions (LHeC4) yield the optimal configuration to constrain $\Clu$ and $\Cld$. This is an interesting result that shows the physics gain resulting from a positron beam at a future LHeC. It arises from the structure of the underlying matrix elements. We discuss this point in more detail later.
    
    \item $e^+p$ collisions with unpolarized positrons (LHeC7 and FCCeh3) serve as the optimal configuration to constrain $\Cqe$.

\end{itemize}

\begin{table}
    [htbp]
    \centering
    \caption{Individual and marginalized 95\% CL bounds on semi-leptonic four-fermion Wilson coefficients at $\Lambda = 1~\TeV$ with the individual EIC data sets P4 and $\Delta$P4, the combined EIC fit of D4, $\Delta$D4, P4, and $\Delta$P4, the individual LHeC runs, the joint LHeC run, the individual FCC-eh runs as well as the joint FCC-eh fit, as well as the corresponding effective UV scales in units of TeV.}
    \label{tab:1d-bounds-uv-table}
    \includegraphics[width=\textwidth]{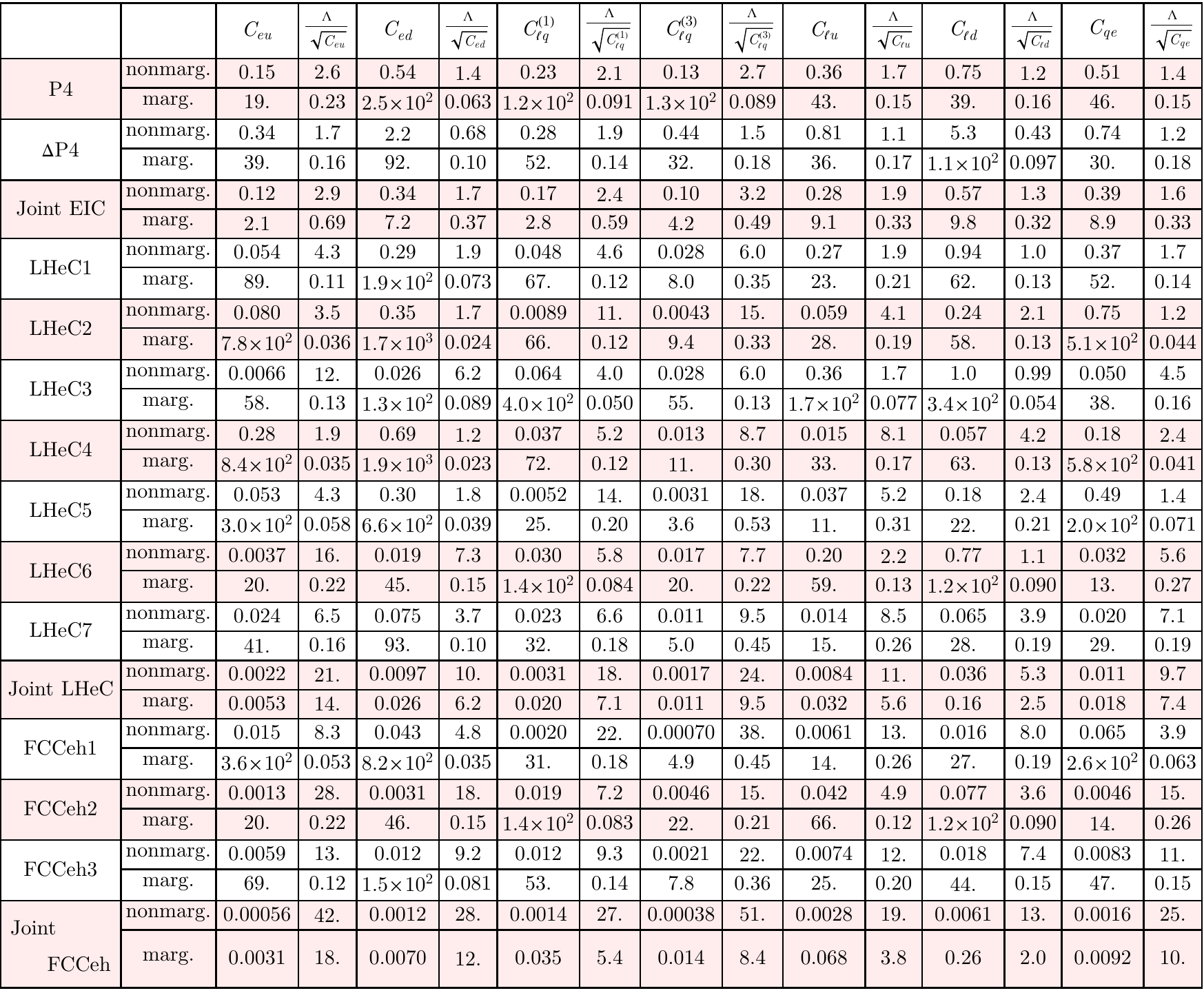}
\end{table}

The effective UV scales presented in Table~\ref{tab:1d-bounds-uv-table} are defined as $\Lambda/\sqrt{C_k}$ for each Wilson coefficient $C_k$.  We note that the convergence of the EFT expansion is controlled by the ratio $C_k Q^2/\Lambda^2$, where $Q$ denotes the DIS momentum transfer. The effective scale constraints obtained above indicate that this ratio is significantly less than unity for all runs considered. This supports our truncation of the expansion at dimension-6, as well as our linearization of the dimension-6 SMEFT effects.

\begin{figure}
    [htbp]
    \centering
    \includegraphics[height=.3\textheight]{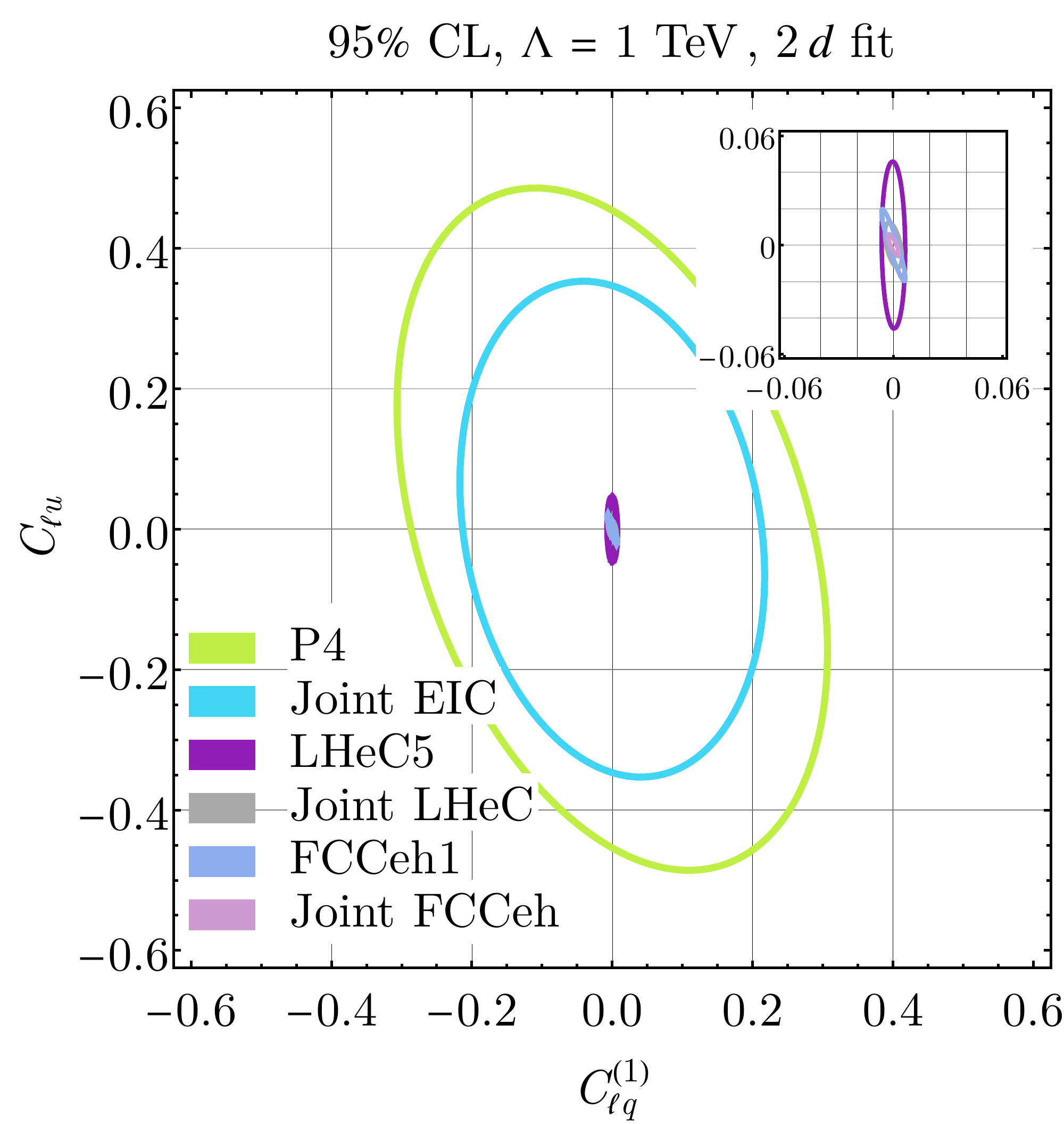}
    \includegraphics[height=.3\textheight]{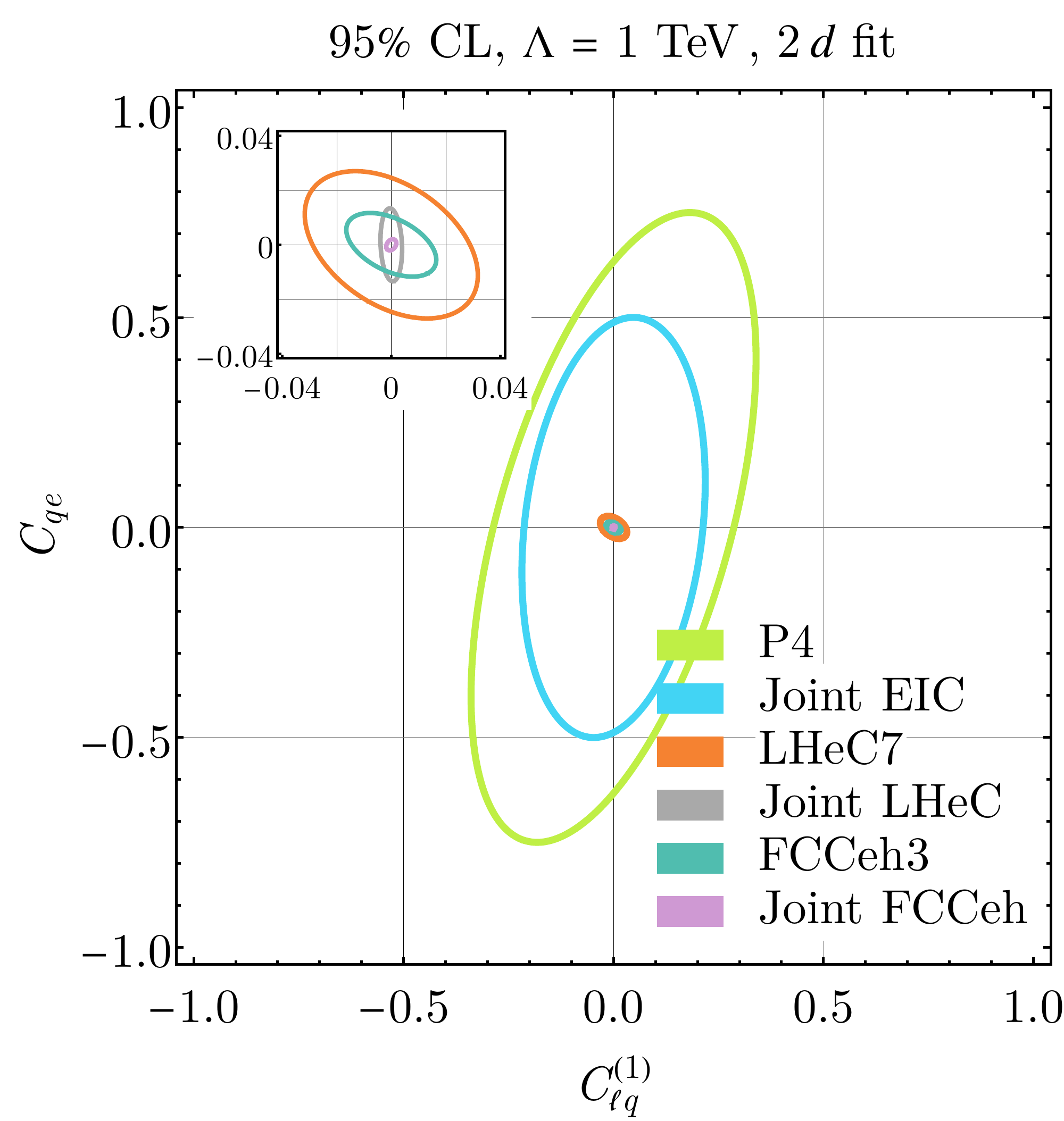}
    \includegraphics[height=.3\textheight]{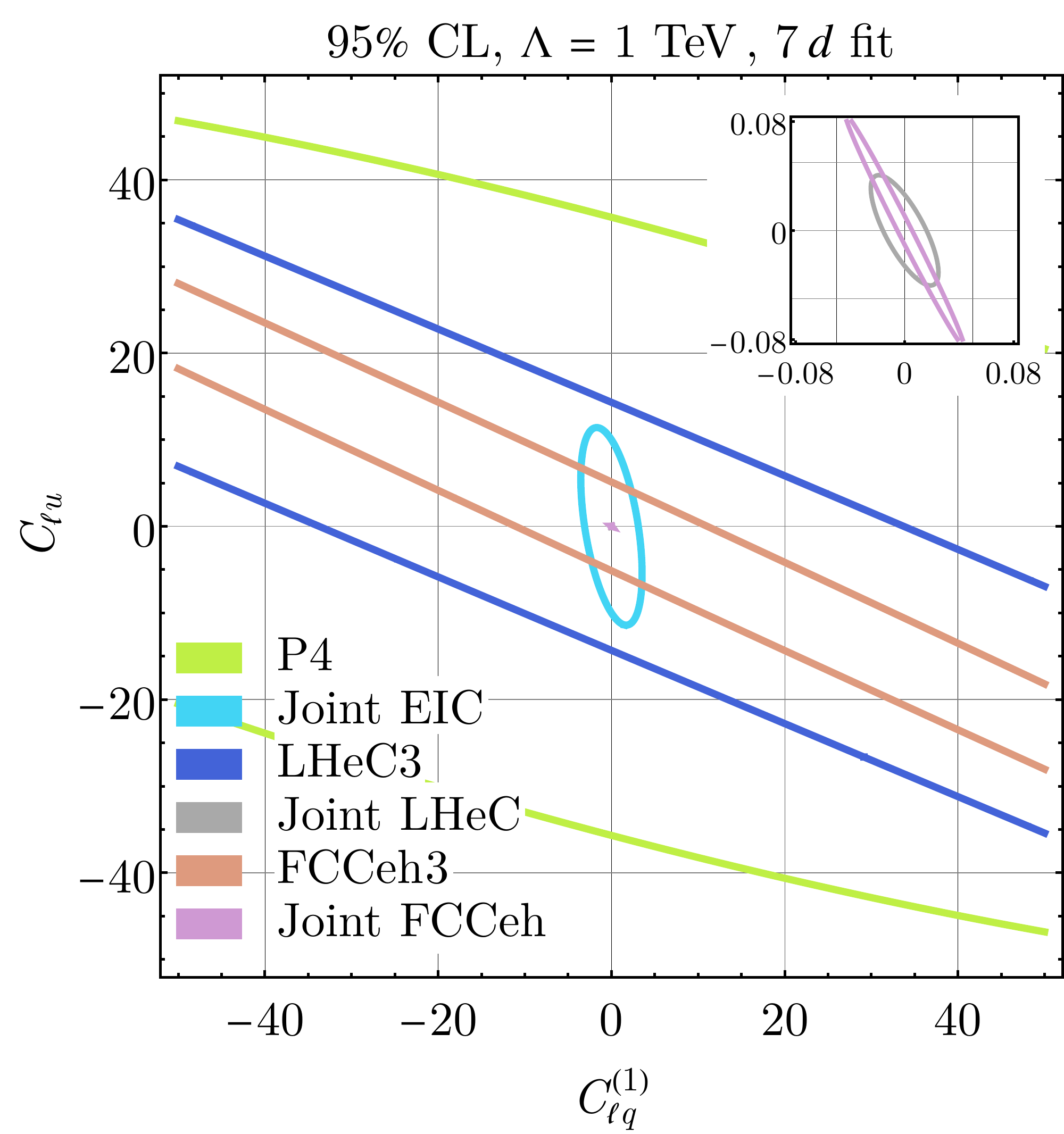}
    \includegraphics[height=.3\textheight]{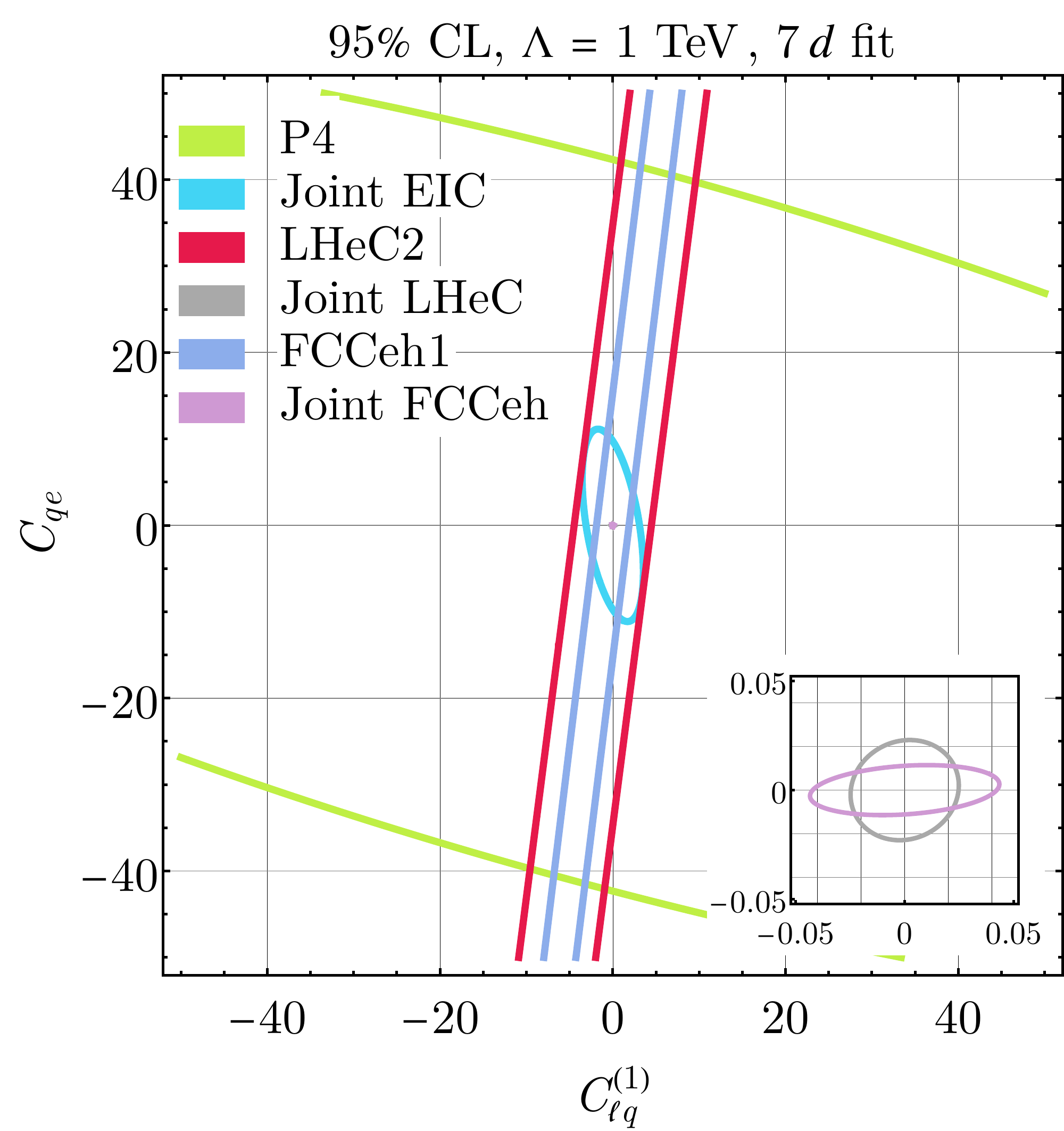}
    \caption{Non-marginalized (top) and marginalized (bottom) 95\% confidence level ellipses for the parameter subspaces spanned by $\Clqone$ and $\Clu$ (left) and $\Clqone$ and $\Cqe$ (right) with $\Lambda = 1~\TeV$. Shown are the strongest individual EIC data set, the strongest LHeC and FCC-eh sets for these Wilson coefficients, as well as the joint EIC, FCC-eh and LHeC fits. The insets show the zoomed-in plot of the joint LHeC and FCC-eh fits.}
    \label{fig:clq1-clu}
\end{figure}

In Fig.~\ref{fig:clq1-clu}, we present representative confidence ellipses projected from the $7d$ fit of the four-fermion Wilson coefficients. In order to emphasize the changes in higher-dimensional fits as more Wilson coefficients are activated, we also include ellipses where only two Wilson coefficients are activated at a time. We show the results for the strongest LHeC, FCC-eh and EIC data sets, as well as the joint fits, for the shown pairs of Wilson coefficients. We present zoomed-in ellipses of the joint FCC-eh and LHeC fits for clarity. Flat directions not present in the 2d fits emerge when all several Wilson coefficients are activated, significantly weakening the bounds obtained from individual run scenarios. However, they are ameliorated in the joint fits, and in particular the LHeC and FCC-eh joint fits show very similar constraints in both the marginalized and non-marginalized cases. The joint EIC constraint ellipse remains weaker in the joint fit, although it is much stronger than the individual P4 fit, indicating the need to consider multiple run scenarios at the EIC.

\begin{figure}
    [htbp]
    \centering
    \includegraphics[height=.3\textheight]{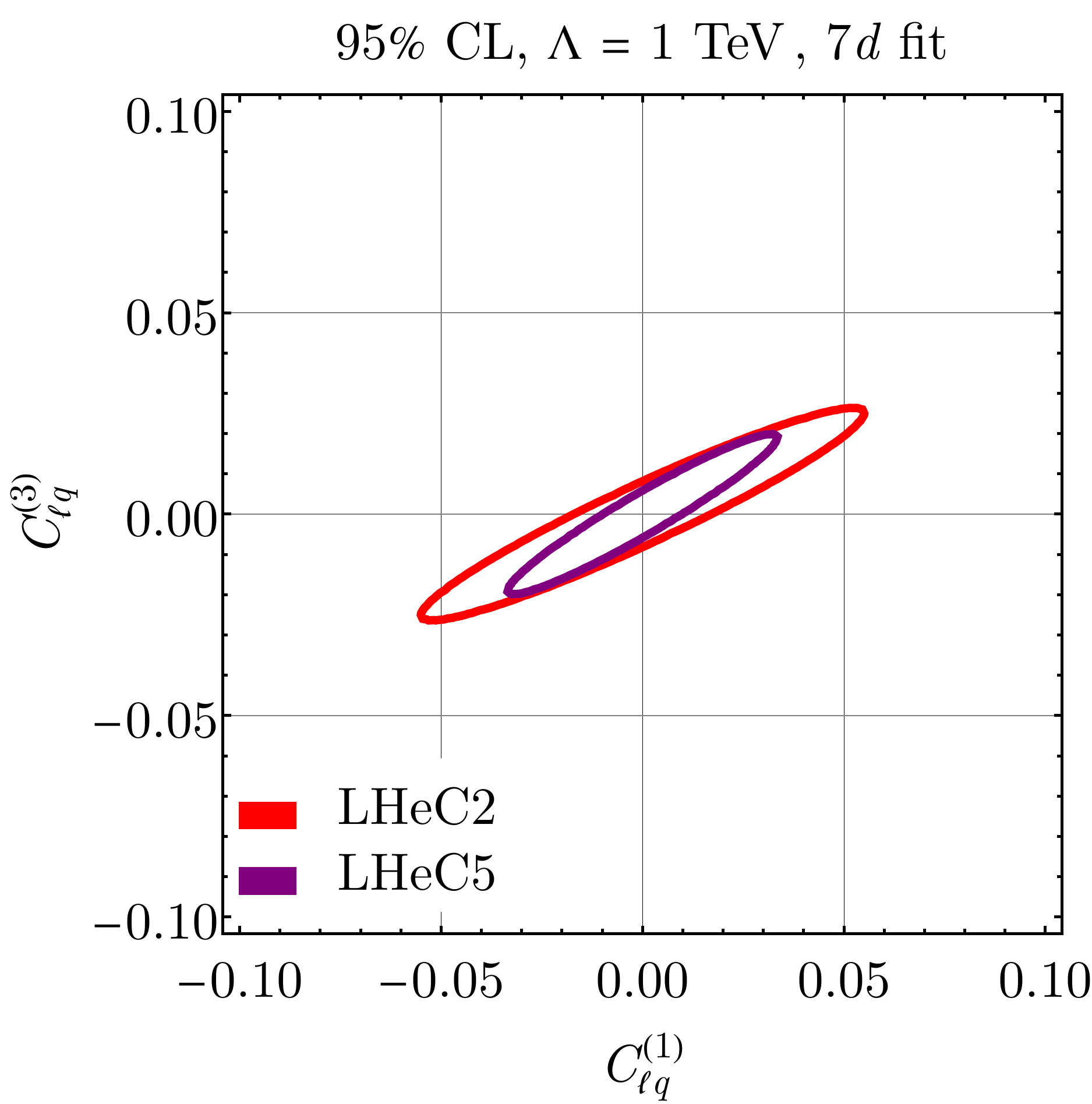}
    \includegraphics[height=.3\textheight]{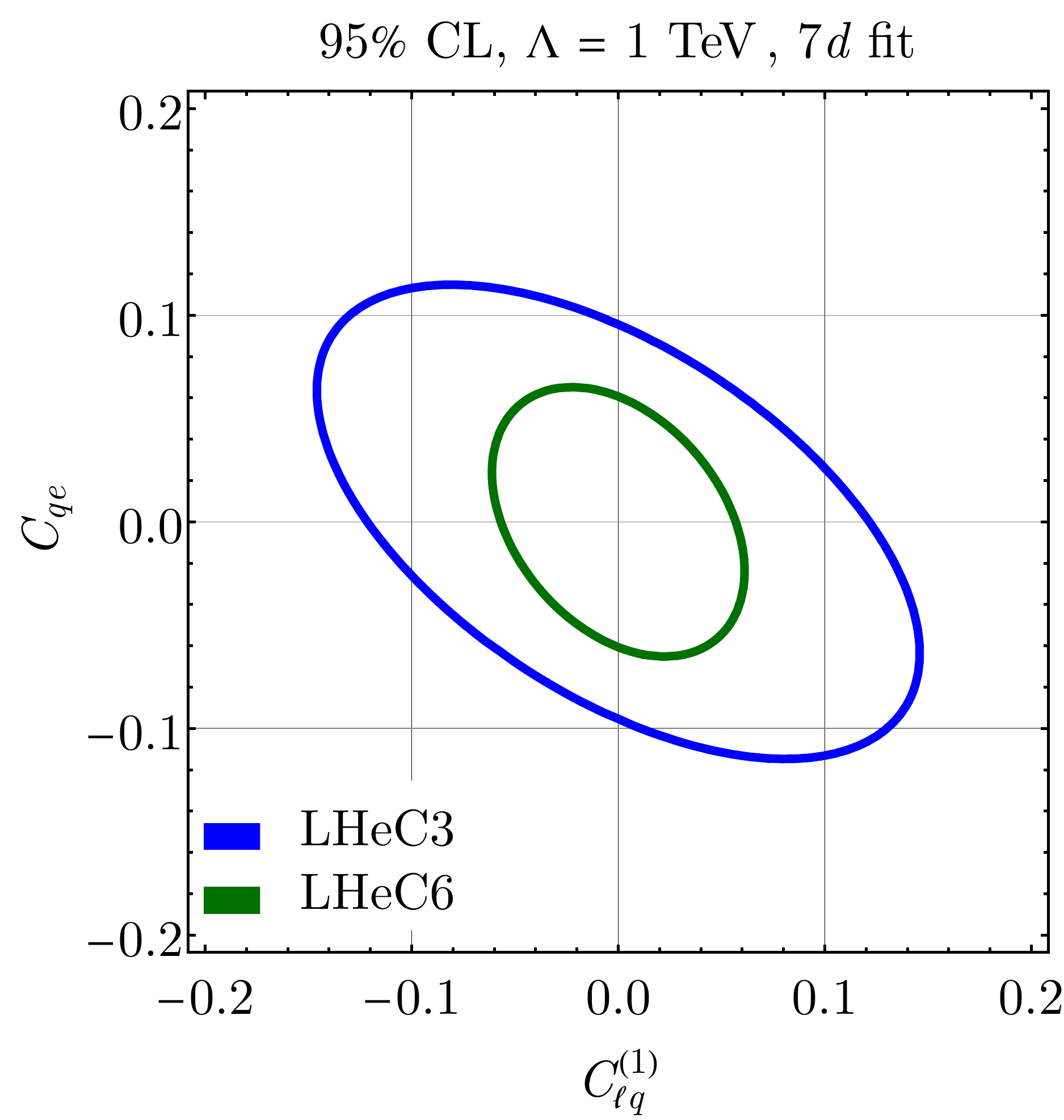}
    \caption{Marginalized 95\% CL ellipses in the parameter subspaces spanned by $\Clqone$ and $\Clqthree$ (left) and $\Clu$ and $\Cld$ (right) at $\Lambda=1~\TeV$ comparing data sets having luminosities that differ by a factor of 10, namely LHeC2 and LHeC5 (left) and LHeC3 and LHeC6 (right).}
    \label{fig:2d-fits-lum-comparison}
\end{figure}

Referring to Table~\ref{tab:data-sets} we note that there are three parameters in the simulations that can be varied: luminosity, lepton beam polarization, and lepton species. We further investigate the physics impact of varying these parameters. For simplicity we will focus this study on the LHeC, although the conclusions hold for the FCC-eh as well. We can compare LHeC2 to LHeC5, and LHeC3 to LHeC6, to check the importance of integrated luminosity. We can also compare LHeC3 and LHeC4 to understand the consequences of having different lepton species. In Fig.~\ref{fig:2d-fits-lum-comparison}, we present plots that compare the impact of increasing the integrated luminosity. Increasing the luminosity (going from LHeC2 to LHeC5 or LHeC3 to LHeC6) only slightly improves the estimated bounds. As shown in Fig.~\ref{fig:lhec-p4-uncertainties} systematic uncertainties dominate over statistical uncertainties at both the LHeC and the FCC-eh, and must be brought under control to facilitate high-luminosity BSM analyses. 

\begin{figure}
    [htbp]
    \centering
    \includegraphics[height=.3\textheight]{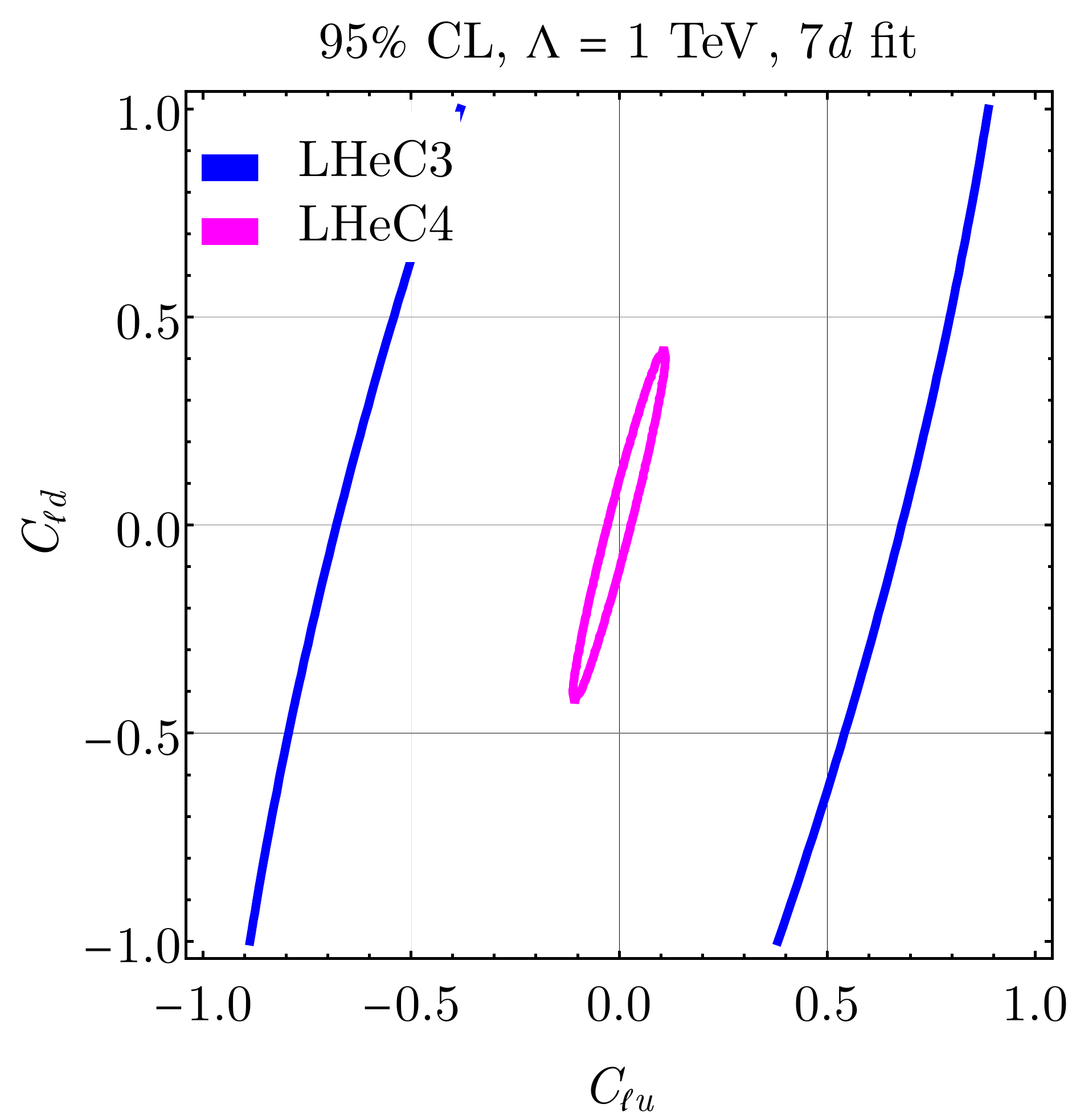}
    \caption{The same as in Figure \ref{fig:2d-fits-lum-comparison} but for $\Clu$ and $\Cld$, with data sets having different lepton species (LHeC3 and LHeC4).}
    \label{fig:2d-fits-lep-comparison}
\end{figure}

Fig.~\ref{fig:2d-fits-lep-comparison} demonstrates that significant improvements occur when we change the lepton species from electrons to positrons. This is despite LHeC4, with positrons, having three times less luminosity than LHeC3 with electrons, as we recall from Table~\ref{tab:data-sets}. The reason is that the $y$-dependence of the various Wilson coefficient structures in the matrix elements changes when we switch from electrons to positrons. We note that replacing an electron with a positron amounts to the following interchanges in the matrix elements: $\Ceu$ by $\Clu$, $\Ced$ by $\Cld$, and $\Clqone \mp \Clqthree$ by $\Cqe$. Referring to Eq.~(7) in Ref.~\cite{Boughezal:2020uwq}, we see that these replacements remove the $(1-y)^2$ factors multiplying the Wilson coefficients. These factors reduce the cross sections for the electron case, since on average $(1-y)^2 \sim 1/4$. Removing them leads to larger corrections from the SMEFT for the positron-induced cross sections. This result demonstrates the usefulness of positron runs in the future DIS program.

Finally, it is known that the LHC has blind spots in the Wilson coefficient parameter space due to the observables measured in the Drell-Yan process, and that measurements of the DIS process can help remove these degeneracies~\cite{Boughezal:2020uwq,Boughezal:2022pmb}. It is possible that the DIS measurements themselves exhibit approximate degeneracies as well. We study that issue here, again focusing on the LHeC for illustrative purposes. Looking at the leading-order matrix elements, we see that there are two kinematic structures, a term proportional to $(1-y)^2$ and a term without $y$ dependence. We must separately set these to zero, for both the up-quark and down-quark channels. Doing so leads to four conditions on the seven semi-leptonic four-fermion Wilson coefficients. We choose to keep
$(\Clqone, \Clqthree, \Cqe)$ as our basis. This leads to the following conditions on the other four Wilson coefficients in order to have all SMEFT-induced corrections to $e^-p$ scattering vanish:
\begin{align}
    \Ceu &= 
        {P_\ell - 1 \over P_\ell + 1} 
        {Q_u - g_+^e g_+^u \hat\eta_{\gamma Z} \over Q_u - g_-^e g_-^u \hat\eta_{\gamma Z}} 
        (\Clqone - \Clqthree) \label{basis1-1}\\
    \Clu &= 
        {P_\ell + 1 \over P_\ell - 1}
        {Q_u - g_-^e g_+^u \hat\eta_{\gamma Z} \over Q_u - g_+^e g_-^u \hat\eta_{\gamma Z}}
        \Cqe \label{basis1-2}\\ 
    \Ced &= 
        {P_\ell - 1 \over P_\ell + 1} 
        {Q_d - g_+^e g_+^d \hat\eta_{\gamma Z} \over Q_d - g_-^e g_-^d \hat\eta_{\gamma Z}} 
        (\Clqone + \Clqthree) \label{basis1-3} \\
    \Cld &= 
        {P_\ell + 1 \over P_\ell - 1}
        {Q_d - g_-^e g_+^d \hat\eta_{\gamma Z} \over Q_d - g_+^e g_-^d \hat\eta_{\gamma Z}}
        \Cqe \label{basis1-4}.
\end{align}
We note that this also removes SMEFT corrections to $e^+p$ scattering upon taking $P_\ell \to -P_\ell$. Here, $Q_{u/d}$ is the up/down quark electric charge, $g_\pm^f = g_V^f \pm g_A^f$, $g_{V/A}^f$ are the usual SM vector/axial fermion couplings to the $Z$ boson, and the energy-dependent $\eta$ factor is defined by
\begin{align}
    \hat\eta_{\gamma Z} = {G_F M_Z^2 \over 2 \sqrt2 \pi \alpha} {Q^2 \over Q^2 + M_Z^2}.
\end{align}

\begin{figure}
    [htbp]
    \centering
    \includegraphics[width=\textwidth]{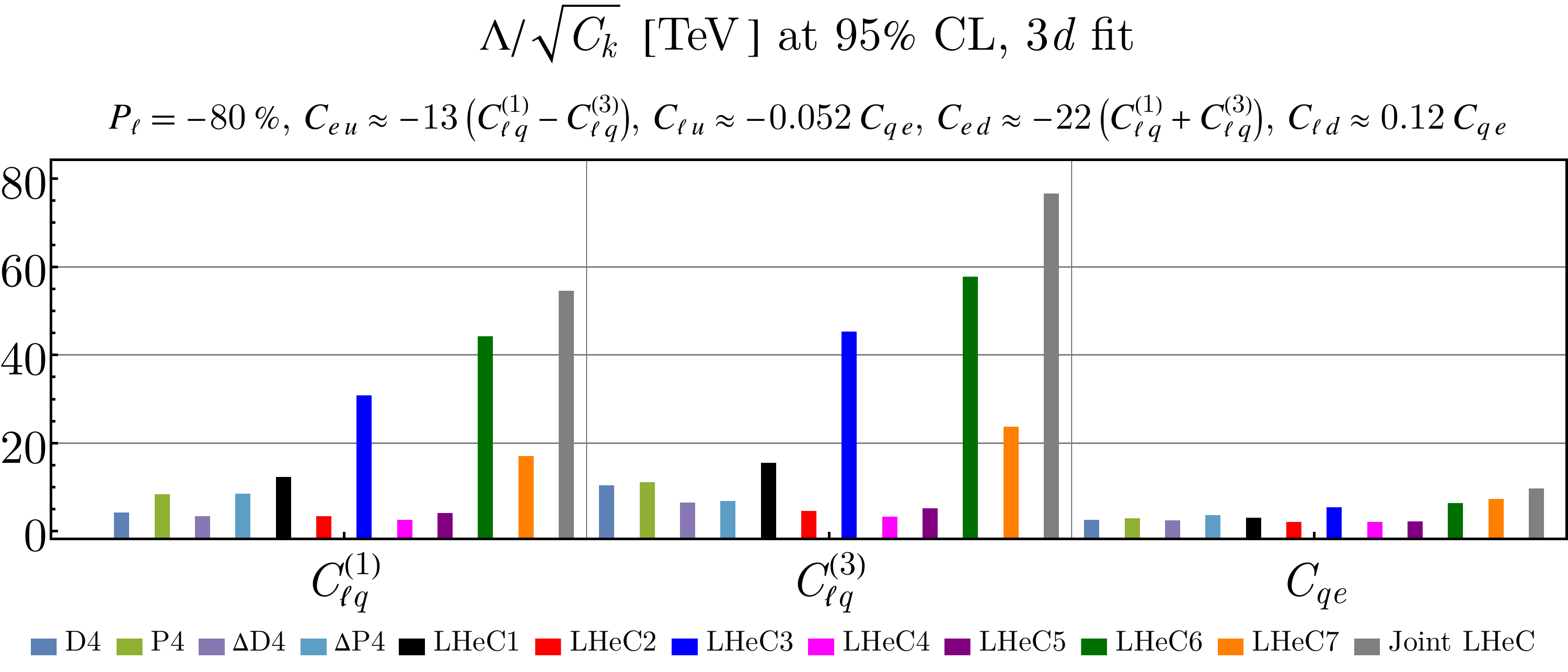}
    \caption{Effective UV scales corresponding to marginalized 95\% CL bounds on the Wilson coefficients $\Clqone$, $\Clqthree$, and $\Cqe$ in the analysis of flat directions for LHeC2, LHeC4, and LHeC5.}
    \label{fig:fd-basis1-case2}
\end{figure}

Before presenting results we first discuss several caveats associated with these solutions. First, due to the presence of the energy-dependent $\eta$ factors, any flat direction can only be approximate. As noted in~\cite{Boughezal:2020uwq} these degeneracies become more apparent at high energies and momentum transfers. Since $Q^2$ reaches up to 1 TeV at the LHeC, and consequently $Q^2 \gg M_Z^2$, we expect them to become important at this experiment. Second, since the solutions above depend on the lepton polarization, a clear path to removing any degeneracy is clear: run the LHeC and FCC-eh with multiple polarization scenarios. This again illustrates the importance of running with multiple run scenarios as outlined in Table~\ref{tab:data-sets}. We also note that this example is a bottom-up construction of a flat direction only, and we make no attempt to connect this to an ultraviolet model. Setting $P_\ell = -80\%$, we can study the approximate flat directions that appear in the fits of LHeC2, LHeC4, and LHeC5 as a representative example. Letting $Q^2/M_Z^2 \to \infty$, Eqs.\eqref{basis1-1} through~\eqref{basis1-4} give
\begin{align}
    \Ceu &\approx -13 (\Clqone - \Clqthree) \equiv \Ceu^{(1)}\\ 
    \Clu &\approx -0.052 \Cqe \equiv \Clu^{(1)} \\ 
    \Ced &\approx -22 (\Clqone + \Clqthree) \equiv \Ced^{(1)} \\ 
    \Cld &\approx 0.12 \Cqe \equiv \Cld^{(1)} .
\end{align}
We now impose these relations and perform fits in the 3-d parameter space of of $\Clqone$, $\Clqthree$, and $\Cqe$. In Fig.~\ref{fig:fd-basis1-case2}, we present effective UV scales derived from the marginalized 95\% CL bounds on $\Clqone$, $\Clqthree$, and $\Cqe$. This figure shows that the reaches of the LHeC2, LHeC4, and LHeC5 runs become weak, as expected. The joint LHeC fit can, however, cover this region of parameter space, as can the EIC. This explicitly demonstrates the importance of running future DIS experiments with multiple parameter scenarios.

\subsection{Bounds on $ffV$ vertex corrections}

We now activate all 17 Wilson coefficients listed in Table \ref{tab:ops}. These include both four-fermion interactions and operators which shift the $ffV$ vertices. One typically expects corrections to the $ffV$ vertices to be better constrained by precision $Z$-pole observables. Indeed, fits with only a single Wilson coefficient activated result in extremely strong bounds, reaching 10 TeV in some cases~\cite{Dawson:2019clf}. However, due to the limited number of measurements possible there are numerous degeneracies in this parameter space. This is nicely illustrated in~\cite{Ellis:2020unq}, where bounds on $ffV$ vertex corrections are loosened by roughly an order of magnitude when switching from single-coefficient fits to results where the other Wilson coefficients are marginalized over. For example, the bound on the effective UV scale associated with the coefficient $C_{\phi WB}$ decreases from approximately 15 TeV to 1 TeV when all coefficients are turned on~(see~\cite{Ellis:2020unq}, Fig.~3). Other possibilities for probing these couplings include top, Higgs and diboson data at the LHC, which are also considered in~\cite{Ellis:2020unq}, and on-shell $Z$-boson production at the LHC~\cite{Breso-Pla:2021qoe}. We consider here the potential of future DIS experiments to probe this sector of the SMEFT.

\begin{figure}
    [htbp]
    \centering
    \includegraphics[width=.6\textwidth]{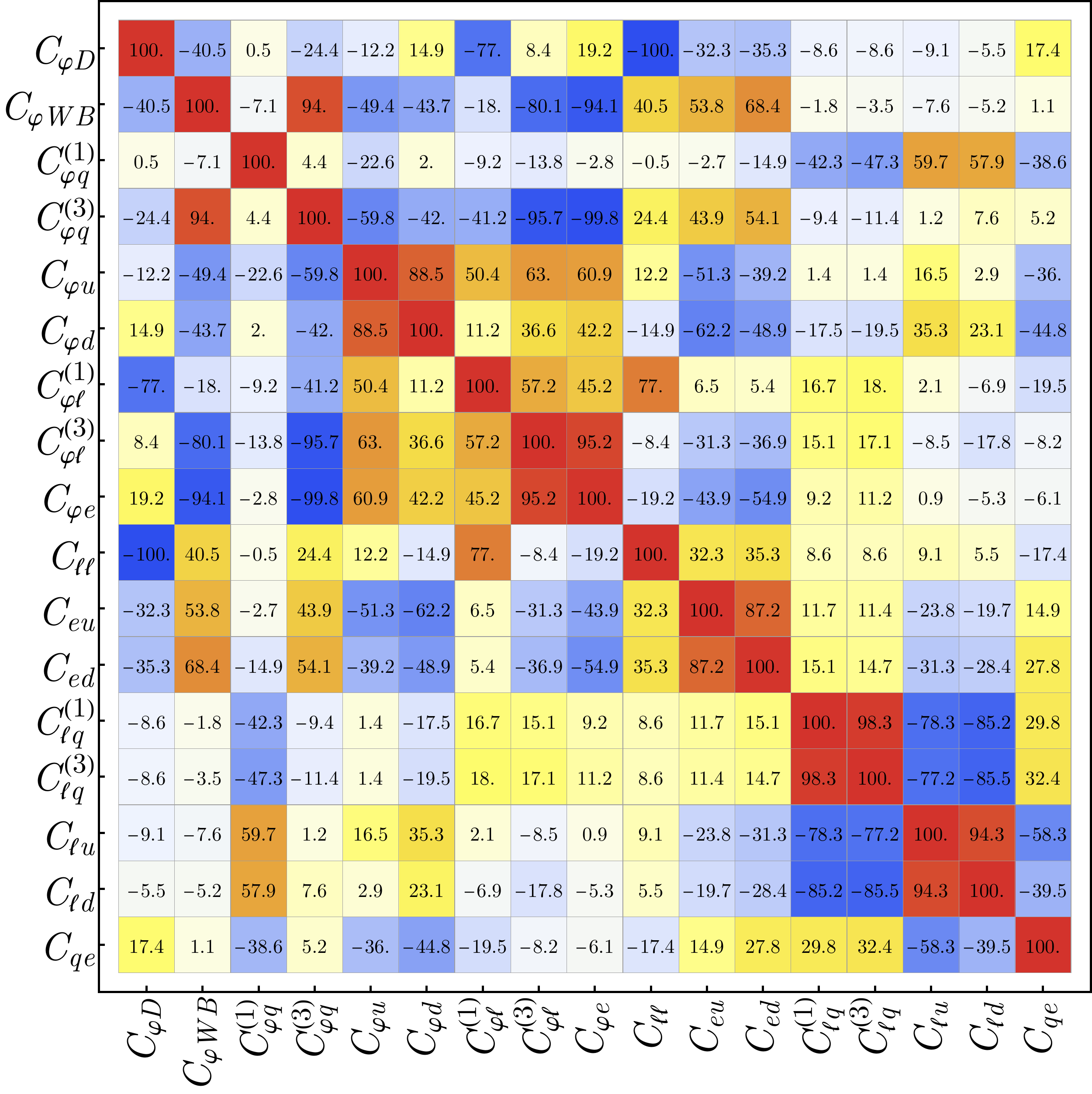}
    \caption{Correlation matrix of the 17d joint LHeC fit of Wilson coefficients.}
    \label{fig:17d-cor-mat}
\end{figure}

\begin{figure}
    [htbp]
    \centering
    \includegraphics[width=.6\textwidth]{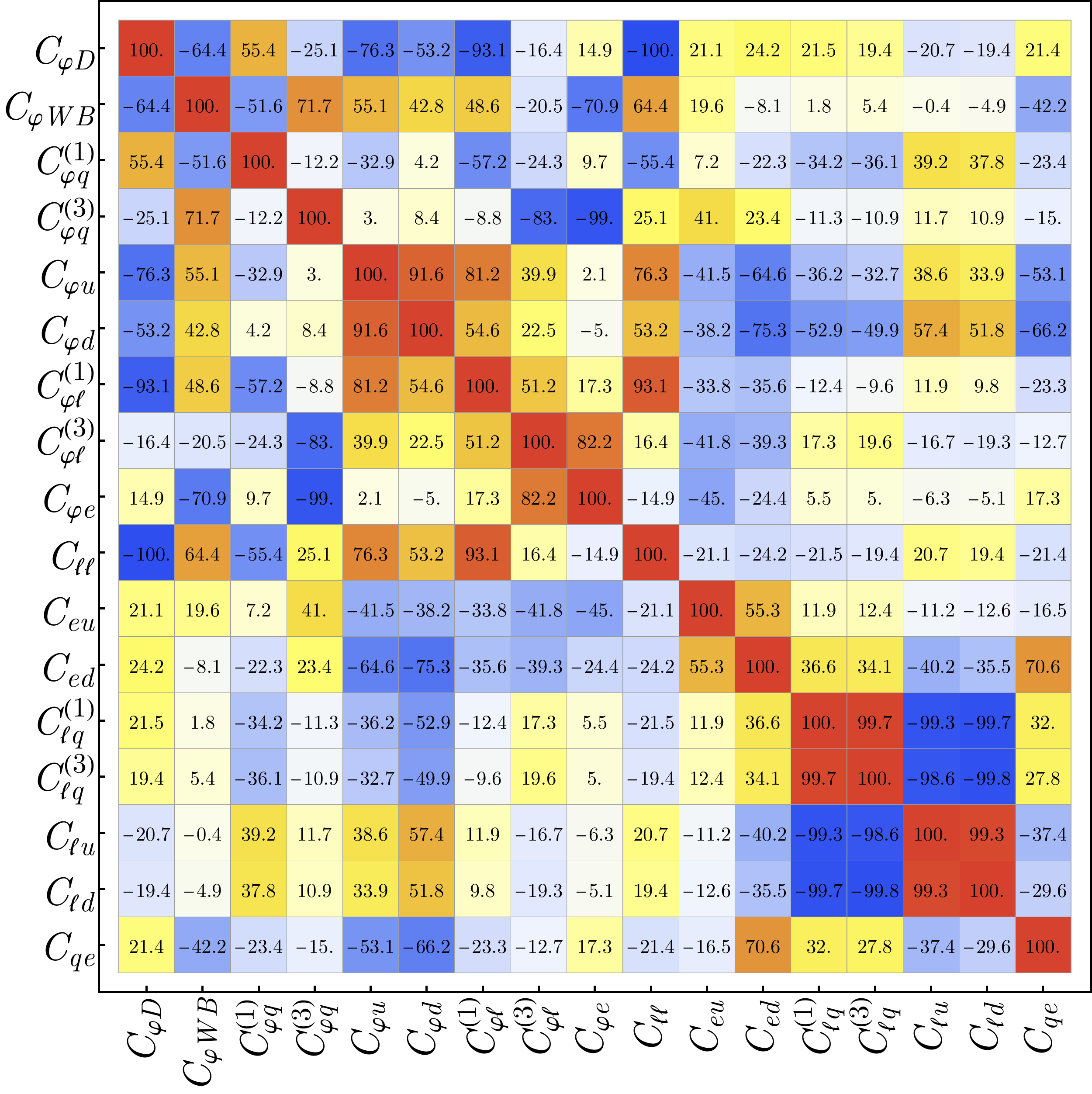}
    \caption{Correlation matrix of the 17d joint FCC-eh fit of Wilson coefficients.}
    \label{fig:17d-cor-mat-fcceh}
\end{figure}

We present in Table~\ref{tab:17d-marg-fits} the marginalized 95\% CL bounds on Wilson coefficients coming from the full 17d fit. We show results for joint EIC fit of D4, $\Delta$D4, P4 and $\Delta$P4, as well as the joint LHeC and FCC-eh constraints. In addition we show the results from the 34d fit of $Z$ and $W$ observables and of EW, diboson, Higgs, and top data, adapted from~\cite{Ellis:2020unq}. To convert the results of~\cite{Ellis:2020unq} to our notation we take the individual non-marginalized 95\% CL bounds, symmetrize them, form pairwise covariance matrices using the given correlations, and then re-derive marginalized 2-parameter fits at 95\% CL. The correlation matrix for our joint LHeC fit is shown in Fig.~\ref{fig:17d-cor-mat}, and the joint FCC-eh fit is shown in Fig.~\ref{fig:17d-cor-mat-fcceh}. We caution that because of the different numbers of parameters fitted in~\cite{Ellis:2020unq}, this is not quite an apples-to-apples comparison between the two fits.

\begin{itemize}

\item The LHeC bounds are stronger than those from the joint fit of electroweak precision data and LHC results for the majority of Wilson coefficients, indicating that it would add constraining power to the global fit. The FCC-eh are stronger than both the LHeC and EIC in most cases.

\item From the correlation matrix we observe that there is only weak correlation between the vertex corrections and the four-fermion operators in the joint LHeC and FCC-eh fits. 

\item The bounds from the EIC reach 500 GeV for the effective scale at most, and are weaker than those obtained from the LHeC and in~\cite{Ellis:2020unq}.

\end{itemize}

\begin{table}
    [htbp]
    \centering
    \caption{Marginalized 95\% CL bounds on Wilson coefficients in the 17d fit assuming $\Lambda = 1~\TeV$, as well as the corresponding effective UV scales in units of TeV. Shown is the combined EIC fit of D4, $\Delta$D4, P4, and $\Delta$P4, the joint LHeC and FCC-eh fits, as well as the marginalized bounds and UV scales from the $34d$ fits of EW, diboson, Higgs, and top data \cite{Ellis:2020unq}.}
    \label{tab:17d-marg-fits}
    \includegraphics[width=.5\textwidth]{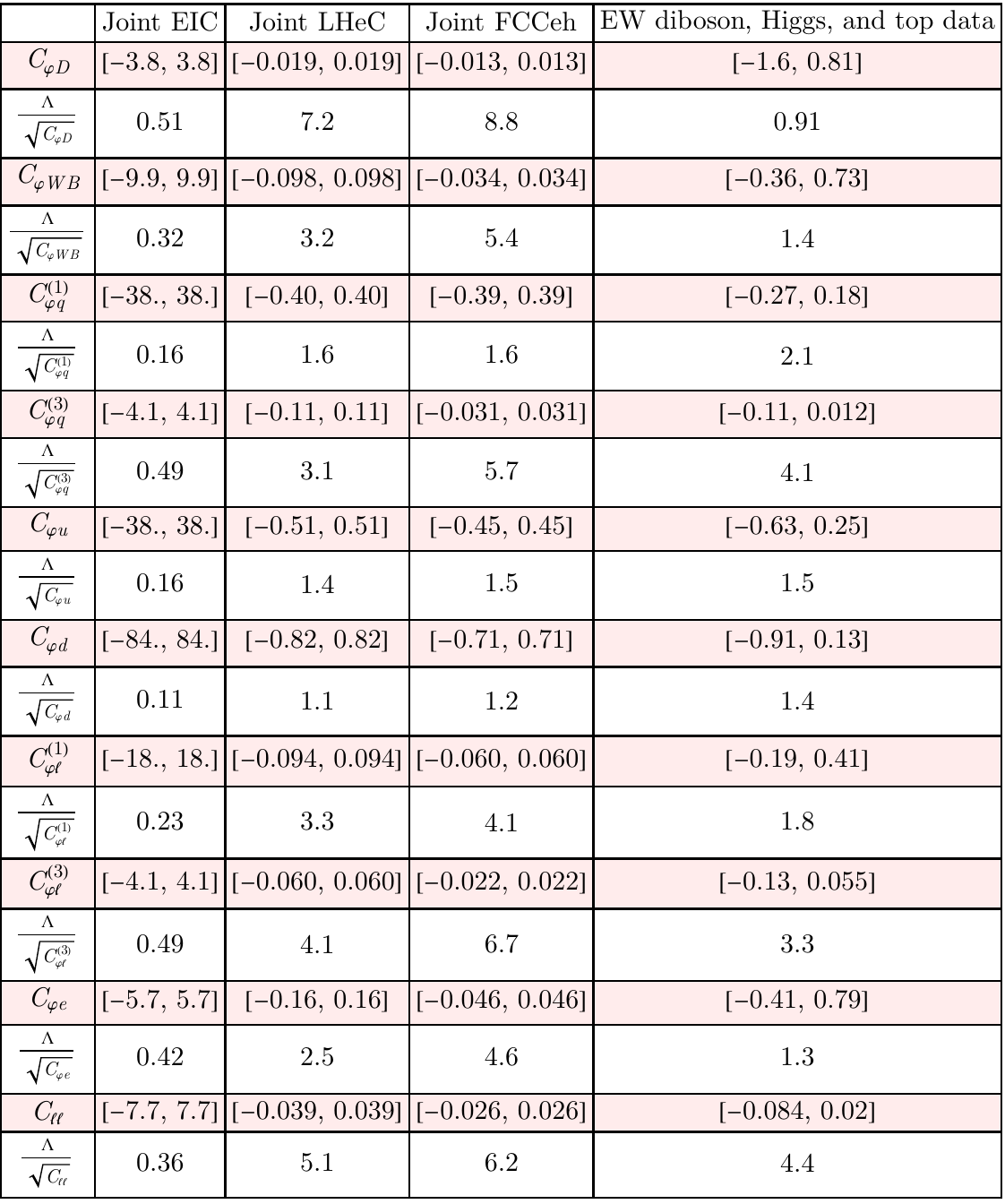}
\end{table}

\begin{figure}
    [htbp]
    \centering
    \includegraphics[height=.3\textheight]{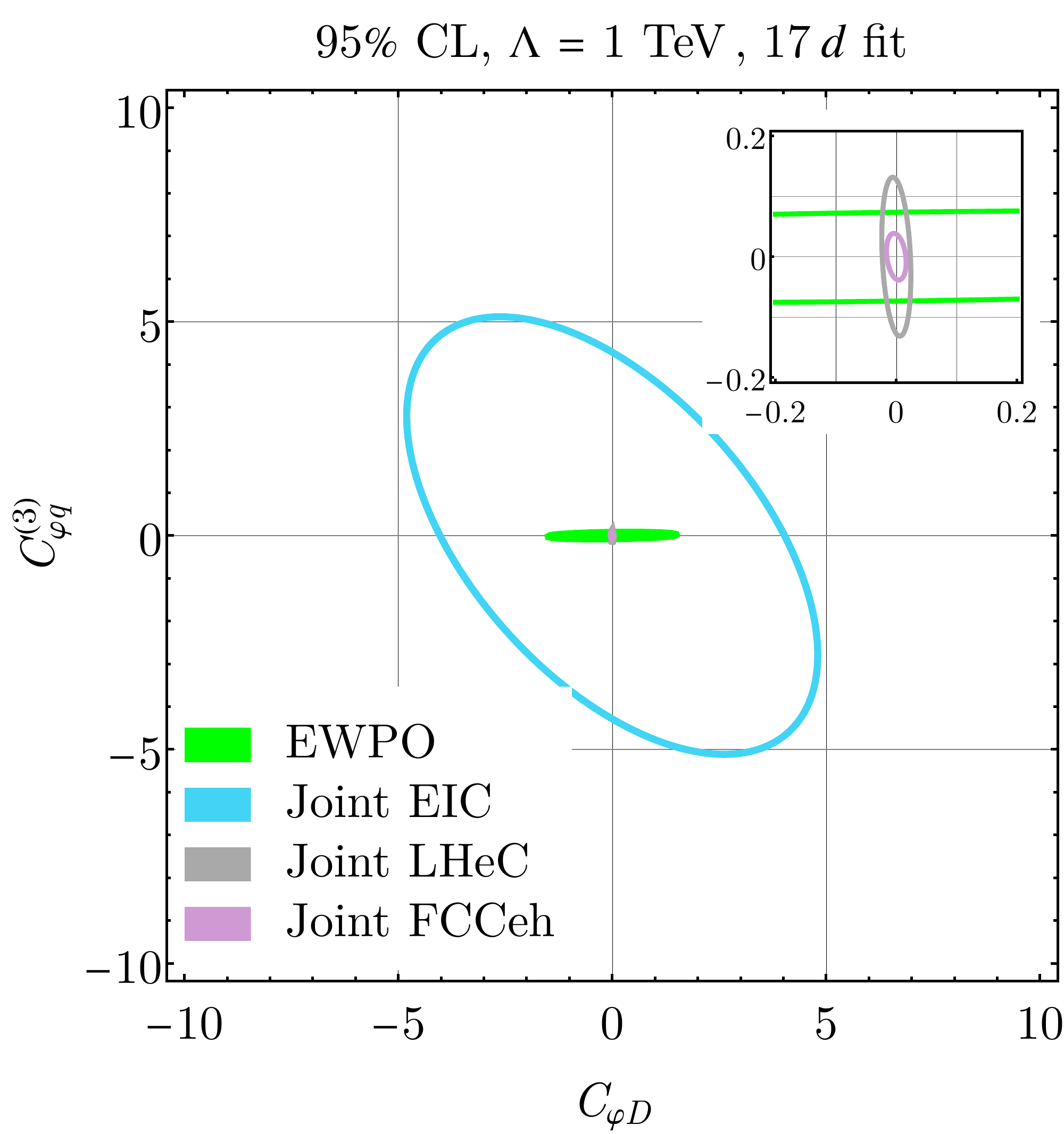}
    \includegraphics[height=.3\textheight]{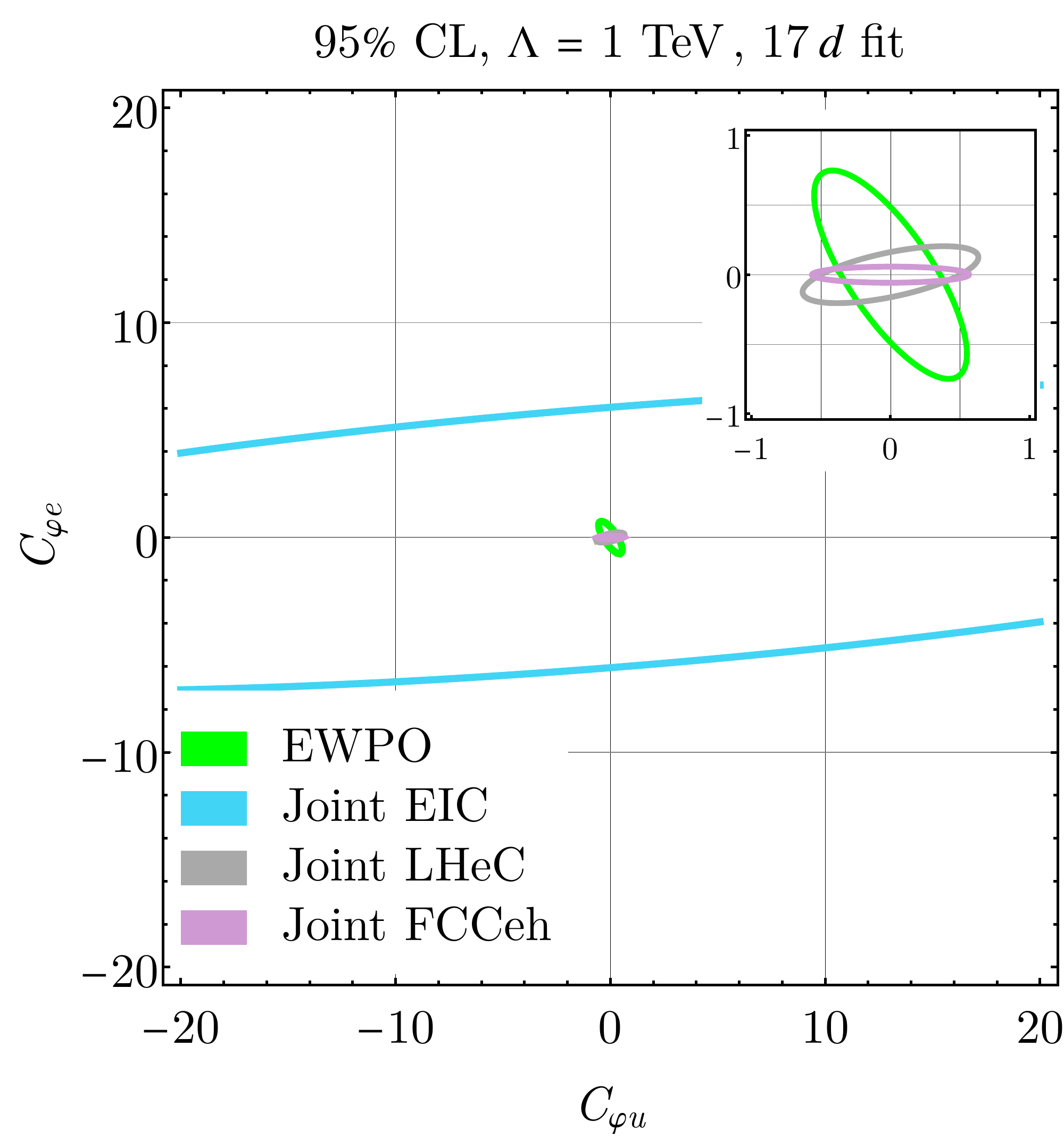}
    \caption{Marginalized 95\% CL ellipses in the two-parameter fits of $C_{\varphi D}$ and $C_{\varphi e}$ (left) and $C_{\varphi\ell}^{(1)}$ and $C_{\varphi e}$ (right) at $\Lambda = 1~\TeV$. Shown are joint EIC, LHeC, and FCC-eh fits, as well as the EWPO fit adapted from \cite{Ellis:2020unq}.}
    \label{fig:ewpo-comparison}
\end{figure}

\begin{figure}
    [htbp]
    \centering
    \includegraphics[height=.3\textheight]{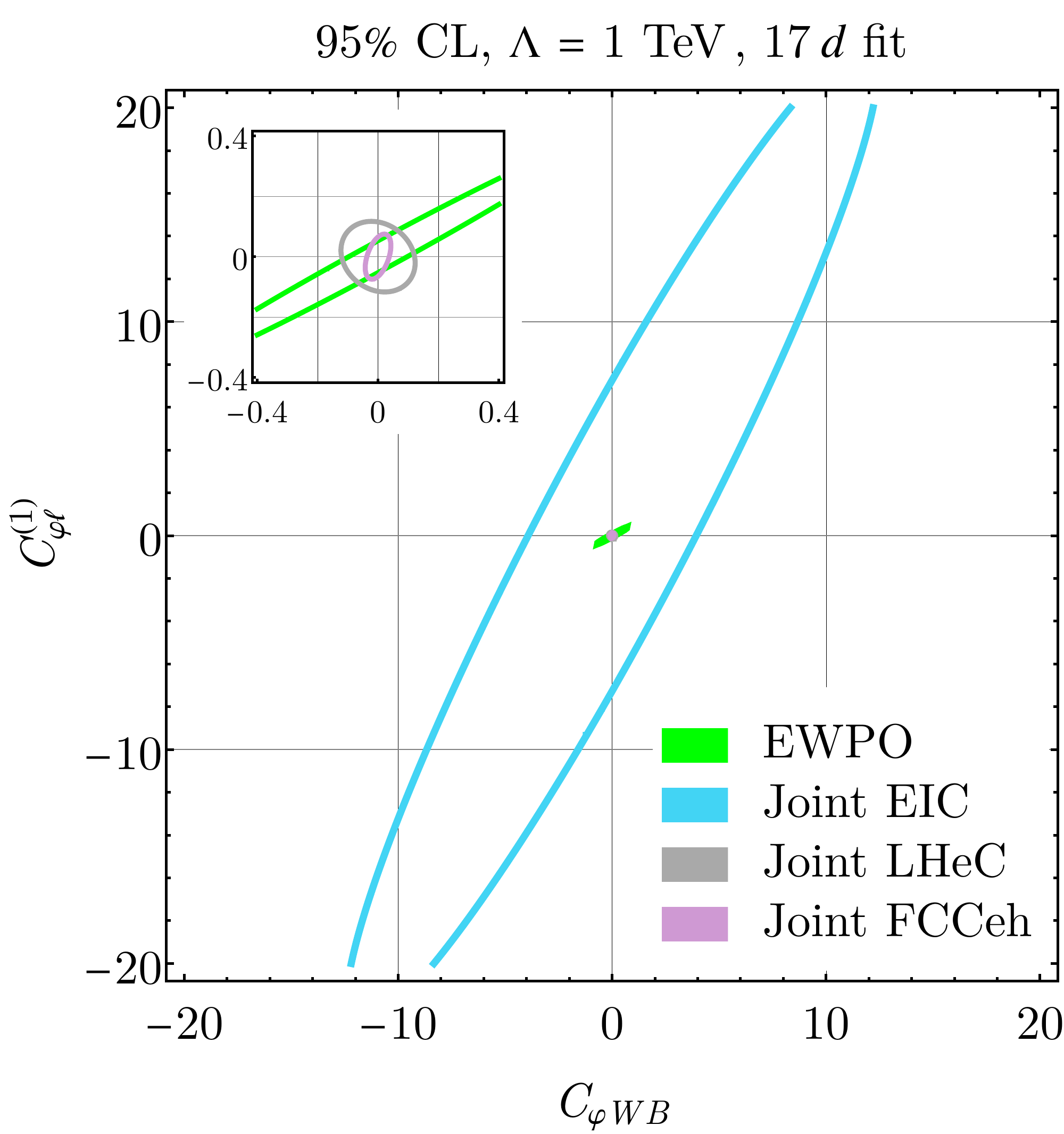}
    \includegraphics[height=.3\textheight]{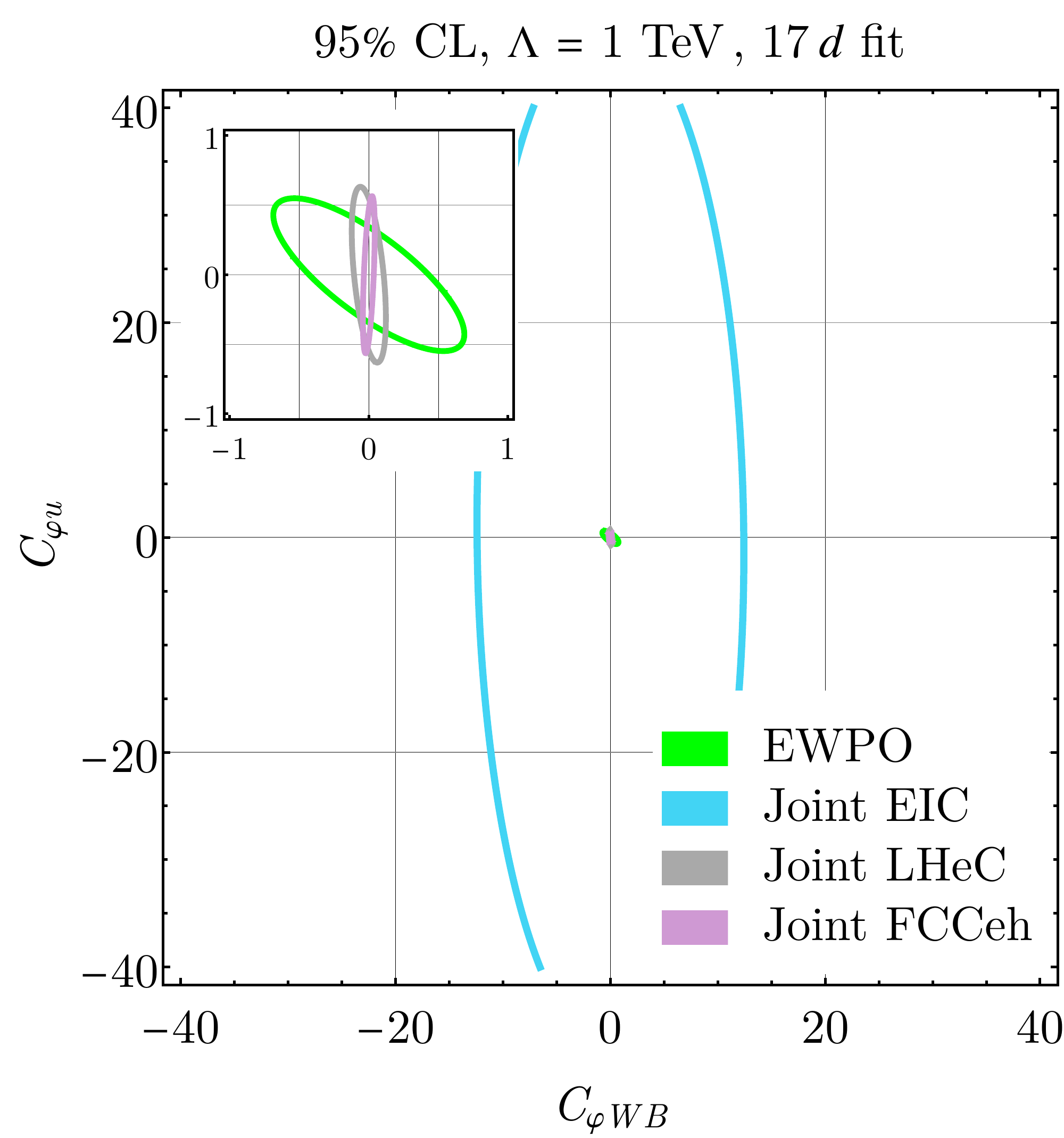}
    \caption{Marginalized 95\% CL ellipses in the two-parameter fits of $C_{\varphi D}$ and $C_{\varphi e}$ (left) and $C_{\varphi\ell}^{(1)}$ and $C_{\varphi e}$ (right) at $\Lambda = 1~\TeV$.  Shown are joint EIC, LHeC, and FCC-eh fits, as well as the EWPO fit adapted from \cite{Ellis:2020unq}.}
    \label{fig:ewpo-comparison2}
\end{figure}

To study this further and to see what including the future precision DIS data in the existing global fit may lead to, we consider several representative 2-d projections of our results. In Figs.~\ref{fig:ewpo-comparison} and~\ref{fig:ewpo-comparison2}, we present non-marginalized 95\% CL ellipses in the parameter subspace spanned by $(C_{\varphi D}, C_{\varphi q}^{(3)})$, $(C_{\varphi u}, C_{\varphi e})$, $(C_{\varphi WB}, C_{\varphi \ell}^{(1)})$, and $(C_{\varphi WB}, C_{\varphi u})$. We consider the joint fits from each DIS experiment, as well as the EWPO fits adapted from \cite{Ellis:2020unq}. We can make the following points from these representative 2d projections.

\begin{itemize}

\item  The potential LHeC probes are in most cases stronger than those of the joint electroweak and LHC fit, and the FCC-eh bounds are stronger still. In particular the joint electroweak and LHC fit exhibits strong correlations between parameters that results in elongated ellipses in several of the 2d projections that we consider, as illustrated by the pairs $(C_{\varphi D}, C_{\varphi q}^{(3)})$ and $(C_{\varphi WB}, C_{\varphi \ell}^{(1)})$. The combinations of future LHeC and FCC-eh runs do not show these correlations, and can remove these approximate degeneracies in the  joint electroweak and LHC fit.

\item The EIC probes are far weaker than those obtained from the other fits, and do not contribute significantly to probing the $ffV$ parameter space.

\end{itemize}

\section{Conclusions\label{sec:conclusion}}
In this work we have studied the BSM potential of the LHeC, FCC-eh and EIC within the SMEFT framework. Following previous studies in the literature the observables considered are the NC DIS cross section at the LHeC and FCC-eh, and parity-violating asymmetries at the EIC. We considered the full spectrum of SMEFT operators that can shift the DIS cross section, including both semi-leptonic four-fermion operators and $ffV$ vertex corrections. This leads us to a 17-dimensional Wilson coefficient parameter space. We considered numerous experimental configurations for these machines, and various energy, polarization and lepton species in order to determine their impact on probes of SMEFT. We have found that the EIC can probe UV scales up to 3 TeV.  This increases to 13 TeV for individual LHeC runs, 14 TeV with the joint LHeC run, and as high as 18 TeV in the joint FCC-eh fit. No single run scenario at any experiment is ideal for probing the full SMEFT parameter space, and for the purpose of BSM studies it will be important to vary polarization and lepton species. Most importantly, we have found that future precision DIS measurements can lift degeneracies present in the precision electroweak fit to $Z$-pole observables. Constraints from the LHeC and FCC-eh are estimated to be in most cases stronger than those coming from combined fits of $Z$-pole and LHC data. Our results further demonstrate  the BSM potential of future DIS studies. 

\vspace{0.5cm}
\noindent
{\bf Acknowledgments:} 
We thank D.~Britzger for suggesting to include an analysis of the FCC-eh capabilities. C.~B. and R.~B. are supported by the DOE contract DE-AC02-06CH11357.  K.~S. is supported by the DOE grant DE-FG02-91ER40684. This research was supported in part through the computational resources and staff contributions provided for the Quest high performance computing facility at Northwestern University which is jointly supported by the Office of the Provost, the Office for Research, and Northwestern University Information Technology.

\appendix

\section{Details of the experimental error matrix and pseudodata generation \label{app:errmat}}
\subsection{Construction of the error matrix}

We discuss here the structure of the error matrix. The experimental error matrix is defined by
\begin{align}
    E_{{\rm exp}, bb'} = \begin{cases}
        (\delta Q_{{\rm unc}, b} \oplus \delta Q_{{\rm cor}, b})^2, & b = b' \\
        \rho_{bb'} \ \delta Q_{{\rm cor}, b} \ \delta Q_{{\rm cor}, b'}, & b \neq b'
    \end{cases}
\end{align}
with $b,b' = \Range(N_{\rm bin})$, where $b$ and $b'$ are the bin indices, $N_{\rm bin}$ is the number of bins, $Q = \sigma_{\rm NC}$ or $(\Delta)A_{\rm PV}$ is the observable, and $\delta Q_{{\rm unc}, b}$ and $\delta Q_{{\rm cor}, b}$ are the uncorrelated and correlated errors summed in quadrature for the $b^{\rm th}$ bin. We define
\begin{align}
    \delta Q_1 \oplus \delta Q_2 \oplus \cdots = \sqrt{\delta Q_1^2 + \delta Q_2^2 + \cdots}
\end{align}
as a shorthand notation. For the correlated errors we assume full correlation between bins: $\rho_{bb'} = 1$. For the LHeC and FCC-eh data sets, we have
\begin{align}
    \delta \sigma_{{\rm unc}, b} &= \delta \sigma_{{\rm stat}, b} \oplus \delta \sigma_{{\rm ueff}, b} ,\\ 
    \delta \sigma_{{\rm cor}, b} &= \delta \sigma_{{\rm sys}, b} ,
\end{align}
with
\begin{align}
    \delta \sigma_{{\rm sys}, b} = \delta \sigma_{{\rm len}, b} \oplus \delta \sigma_{{\rm lpol}, b} \oplus \delta \sigma_{{\rm hen}, b} \oplus \delta \sigma_{{\rm rad}, b} \oplus \delta \sigma_{{\rm gam}, b} \oplus \delta \sigma_{{\rm geff}, b} \oplus  \delta \sigma_{{\rm lum}, b} .\label{lhec-syst}
\end{align}
The meaning of each individual systematic error in Eq.~(\ref{lhec-syst}) was discussed in Section~\ref{sec:pseudodata}. For the EIC data sets we have
\begin{align}
    \delta (\Delta) A_{{\rm PV, unc}, b} &= \delta (\Delta) A_{{\rm PV, stat}, b} \oplus \delta (\Delta) A_{{\rm PV, sys}, b} \\    
    \delta (\Delta) A_{{\rm PV, cor}, b} &= \delta (\Delta) A_{{\rm PV, pol}, b} \\ 
\end{align}
In addition to the experimental errors we must consider the PDF uncertainties. Potential uncertainties from uncalculated higher-order QCD corrections, typically estimated by varying renormalization and factorization scales, are smaller than the other sources of uncertainty and are neglected in our analysis. The PDF error matrix is defined by
\begin{align}
    E_{{\rm pdf}, bb'} = {1 \over N_{\rm pdf}} \sum_{m = 1}^{N_{\rm pdf}} (Q_{m, b} - Q_{0, b}) (Q_{m, b'} - Q_{0, b'})
\end{align}
where $N_{\rm pdf}$ is the number of PDF members and $Q_{0(m), b}$ is the SM prediction for the observable $Q$ at the $b^{\rm th}$ bin, evaluated with the central ($m^{\rm th}$) member of the relevant PDF set. The total error matrix is given by
\begin{align}
    E = E_{\rm exp} + E_{\rm pdf} \label{individual-err-mat}
\end{align}

In our analysis, we also consider joint fits of various data sets. We assume that the PDF errors and all systematic uncertainties, except the photoproduction background, are correlated among runs. The joint error matrix is given by the individual error matrices of the runs on the block-diagonal entries, with error matrices of correlated uncertainties in the off-block diagonal entries given by
\begin{align}
    J_{nn'} = J_{{\rm exp},nn'} + J_{{\rm pdf},nn'}.
\end{align}
Here, $n, n' $ are the run indices and
\begin{align}
    J_{{\rm exp}, nn', bb'} &= \rho_{nn', bb'} \ \widetilde{\delta Q}_{{\rm cor}, n, b}\ \widetilde{\delta Q}_{{\rm cor}, n', b'} \\ 
    J_{{\rm pdf}, nn', bb'} &= {1 \over N_{\rm pdf}} \sum_{m=1}^{N_{\rm pdf}} (Q_{n,m,b} - Q_{n,0,b}) (\sigma_{{\rm NC}, n',m,b'} - Q_{ n',0,b'}).
\end{align}
The index, $b=1,\ldots,N_{{\rm bin},n}$, $N_{{\rm bin}, n}$ denotes the number of bins of the $n^{\rm th}$ data set. $\widetilde{\delta \sigma}_{{\rm cor}, n, b}$ is given by Eq.\eqref{lhec-syst} after removing the photoproduction background error, $\delta\sigma_{{\rm gam}, b}$. $Q_{, n, 0(m), b}$ is the observable evaluated with the central ($m^{\rm th}$) member of the PDF set in the $b^{\rm th}$ bin of the $n^{\rm th}$ run. The joint error matrix takes the form
\begin{align}
    E = \pmat{
        E_1 & J_{12} & \cdots & J_{17} \\ 
            & E_2    & \cdots & J_{27} \\
            &        & \ddots & \vdots \\
            &        &        & E_7
    }_{\rm sym}
\end{align}
where $E_n$ is the error matrix of the $n^{\rm th}$ set given by Eq.~\eqref{individual-err-mat}.

\subsection{Generation of the pseudodata}

Following the procedure of~\cite{Boughezal:2022pmb} we simulate numerous realizations of each LHeC, FCC-eh and EIC run, which we denote as pseudoexperiments.  For each pseudoexperiment we define a $\chi^2$ test function by
\begin{align}
    \chi_e^2 = \sum_{b,b' = 1}^{\rm N_{\rm bin}} (Q_b - Q_{e,b}) \hat E^{-1}_{bb'} (Q_{b'} - Q_{e,b'})
\end{align}
where $Q_b$ is the SMEFT expression and $Q_{e, b}$ is the simulated value for the observable $Q$ in the $b^{\rm th}$ bin. Here, $\hat E^{-1}$ indicates the symmetrized inverse error matrix,
\begin{align}
    \hat E^{-1} = \frac12 [E^{-1} + (E^{-1})^{\rm T}]
\end{align}
For a given observable $Q$ we define pseudoexperimental values according to
\begin{align}
    Q_{e,b} = Q_b^{\rm SM} + r_{e,b} \ \delta Q_{{\rm unc}, b} + \sum_j r_{j,e}' \ \delta Q_{{\rm cor}_j, b}
\end{align}
where $Q_b^{\rm SM}$ is the SM prediction for the observable $Q$. $r_{e,b}$ and $r'_{j,e}$ are random variables picked from the unit normal distribution, namely $r_{e,b}, r'_{j,e} \sim \mathcal N(0, 1)$. $\delta Q_{{\rm unc},b}$ is the total uncorrelated uncertainty, and $\delta Q_{{\rm cor}_j, b}$ is the $j^{\rm th}$ correlated uncertainty. Note that each correlated error is introduced with a single random variable for each pseudoexperiment.

\subsection{Statistical treatment for Wilson coefficient bounds}

The $\chi^2$ function for the joint LHeC and FCC-eh runs, which have 206 and 120 bins respectively, for a single pseudoexperiment has the form
\begin{align}
    \chi_e^2 (\vec r_e, \vec r'_e) = \chi^2_{{\rm SM}, e} (\vec r_e, \vec r'_e) + \vec \omega_e (\vec r_e, \vec r'_e) \cdot \vec C + \vec C \cdot M \vec C
\end{align}
where $\vec r_e$ and $\vec r'_e$ stand for all the random variables involved. The best-fit values of the Wilson coefficients, $\bar{\vec C}_e$, are given by minimizing the $\chi^2_e$ function for each pseudoexperiment via
\begin{align}
    \left.{\del \chi_e^2 \over \del \vec C} \right|_{\vec C = \bar{\vec C}_e} = 0.
\end{align}
The inverse covariance matrix of the fit is obtained from the second derivatives of the $\chi^2_e$ function as 
\begin{align}
    V^{-1} = \frac12 \left. {\del^2 \chi_e^2 \over \del \vec C \ \del \vec C} \right|_{\vec C = \bar{\vec C}_e} = M.
\end{align}
This is constant for all the pseudoexperiments, hence we have dropped the subscript $e$. The average of the best-fit values of Wilson coefficients across pseudoexperiments, $\bar{\vec C}$, is given by
\begin{align}
    \bar{\vec C} = \bb{\sum_{e = 1}^{N_{\rm exp}} V^{-1}}^{-1} \bb{\sum_{e=1}^{N_{\rm exp}} V^{-1} \bar{\vec C_e}} = {1 \over N_{\rm exp}} \sum_{e=1}^{N_{\rm exp}} \bar{\vec C_e}
\end{align}
where $N_{\rm exp}$ is the number of pseudoexperiments. Since the inverse covariance matrix is constant for all pseudoexperiments, we manage to avoid running a large number of pseudoexperiments, which saves a great deal of computational expense. We justify this approach by noting that the distribution of the best-fit values of each Wilson coefficient exhibits a Gaussian distribution around zero. We know that the average of the best-fit values of each Wilson coefficient is expected to be zero, which is in fact the case for large $N_{\rm exp}$. 

The marginalized bound for the Wilson coefficient $C_k$ is $[-\Delta C_k, \Delta C_k]$ where
\begin{align}
    \Delta C_k = \sqrt{\Delta \chi^2(d,c) \over V^{-1}_{kk}}.
\end{align}
The confidence ellipse in the parameter subspace spanned by Wilson coefficients $C_k$ and $C_{k'}$ is described by
\begin{align}
    \pmat{C_k & C_{k'}} V^{-1}_{kk'} \pmat{C_k \\ C_{k'}} = \Delta \chi^2(d,c)
\end{align}
where $\Delta \chi^2(d,c)$ is the quantile of the $\chi^2$ distribution for $d$ fitted parameters at confidence level $c$. Here, $V^{-1}_{kk}$ is the inverse of the $kk$ entry of the covariance matrix, and $V^{-1}_{kk'}$ is the inverse of the covariance matrix after removing all the rows (columns) other than the $k$th ($k'$th) ones.

\bibliography{refs}

\end{document}